\documentclass[prl,superscriptaddress,twocolumn]{revtex4-2}
\usepackage{times,amsmath}
\usepackage{epsfig}
\usepackage{color}
\usepackage{graphicx}
\usepackage{dcolumn}
\usepackage{bm}
\usepackage{bookmark}
\usepackage{tabularx}
\usepackage{hyperref}
\usepackage{multirow}
\usepackage{array,mathtools,amssymb,booktabs,makecell}
\hypersetup{colorlinks=true, citecolor=blue, filecolor=blue, linkcolor=blue, urlcolor=blue}

\usepackage{proofread}

\begin{document}

\title{Orbital Selective Kondo Effect in Heavy Fermion Superconductor UTe$_{2}$}

\author{Byungkyun Kang}\email{byungkyun.kang@unlv.edu}
\affiliation{Department of Physics and Astronomy, University of Nevada, Las Vegas, Nevada 89154, USA}
\affiliation{Condensed Matter Physics and Materials Science Department, Brookhaven National Laboratory, Upton, NY 11973, USA}

\author{Sangkook Choi}
\affiliation{Condensed Matter Physics and Materials Science Department, Brookhaven National Laboratory, Upton, NY 11973, USA}

\author{Hyunsoo Kim}\email{hyunsoo.kim@ttu.edu}
\affiliation{Department of Physics and Astronomy, Texas Tech University, Lubbock, Texas 79409-1051, USA}

\date{\today}

\begin{abstract}
It has been a great challenge to explore many-body effects in heavy fermion systems with $ab$-$initio$ approaches.
We computed the electronic structure of UTe$_{2}$ without purposive judgements, such as intentional selection of on-site Coulomb interaction and disregarding spin-orbit coupling.
We show that U-5$f$ electrons are highly localized in the paramagnetic normal state, giving rise to the Kondo effect.
It is also found that the hybridization between U-5$f$ and U-6$d$ predominantly in the orthorhombic $ab$-plane is responsible for the high-temperature Kondo effect.
In contrast, the hybridization between U-5$f$ and Te-5$p$ along the $c$-axis manifests the Kondo scattering at a much lower temperature, which could be responsible for the low-temperature upturn of the $c$-axis resistivity. 
Our results show that the electron correlation in UTe$_2$ is orbital selective, which naturally elucidates the recent experimental observations of anomalous temperature dependence of resistivity.
Furthermore, we suggest that the Kondo effect is suppressed at high pressure owing to weak localization of magnetic moments, which results from enhanced U-5$f$ electron hopping.
\end{abstract}


\vskip 300 pt

\maketitle

{\it Introduction -}
The uranium-based superconductors are promising candidates for the realization of the spin-triplet superconductivity \cite{Aoki2019review}. 
The prominent examples include URhGe \cite{Aoki2001} and UCoGe \cite{Huy2007} with the critical temperature $T_c=$ 0.25 K and 0.8 K, respectively. 
Both compounds undergo the superconducting phase transition from the ferromagnetically ordered state, making the equal spin pairing plausible \cite{Aoki2019review}. 
Recently, superconductivity was discovered in {\it Curie paramagnetic} UTe$_2$ with $T_c \approx 2$ K \cite{Ran2019science,Rosa2021,Aoki2021review} that is one of the highest among the known uranium-based superconductors. 
$T_c$ can be further enhanced nearly two-fold at approximately 1 GPa \cite{Ran2020,Knafo2021}. 
However, the mechanism behind this celebrated high $T_c$ has not been resolved. 
The main obstacle is insufficient knowledge of the electronic structure of the normal state in UTe$_2$, which is a prerequisite for the understanding of superconductivity.

UTe$_2$ exhibits the iconic incoherent-to-coherent crossover around $T=$ 50 K in the resistivity with an electrical current along the orthorhombic $a$-direction \cite{Ran2019science,Aoki2019}, which is reminiscent of prototypical Kondo lattice YbRh$_2$Si$_2$ \cite{Trovarelli2000}. 
Both UTe$_2$ and YbRh$_2$Si$_2$ exhibit a negative slope in electrical resistivity at room temperature, implying that inelastic scattering dominates over the electron-phonon scattering.
The magnetic contribution of resistivity is suggestive of $-\ln{T}$ dependence observed in most of the Kondo lattice systems above the coherent temperature $T^*$, for instance, in-plane resistivity in CeCoIn$_5$ between $T=40$ K and 180 K \cite{Petrovic2001}. 
While the Kondo scattering is not the only possibility for the resistivity upturn, recent spectroscopy experiments successfully observed the formation of a hybridized band well above the coherent temperature in CeCoIn$_5$ \cite{Aynajian2012, Chen2017}. 
Most notably a combined study of ARPES and DMFT identified the occurrence of Kondo resonance up to 200 K in CeCoIn$_5$ \cite{jang2020evolution}. 
In contrast, no hybridization gap has been observed in UTe$_2$ at any temperature, leaving the scattering mechanism in the normal state elusive in this new heavy-fermion compound.

The $c$-axis transport property in UTe$_2$ is qualitatively different where it is metallic between 50 K  and at least 300 K. 
The $c$-axis resistivity shows a rapid upturn below 50 K before exhibiting coherent-like behavior at the onset around 13 K \cite{Eo2021}. 
Moreover, magnetic susceptibility below $\sim 100$ K shows distinct temperature-dependence in all three symmetry directions \cite{Ran2019science,Aoki2019}. 
The anisotropic transport and magnetic properties suggest that the orbital-dependent electron correlation needs to be taken into account.
These outstanding issues put heavy-fermion UTe$_2$ in a league of its own, and thus the understanding of the electronic structure of its normal state is the most pressing issue. 
While several ARPES studies were reported \cite{Fujimori2019,Miao2020}, a complete picture of the band structure is still awaited. 
Therefore, theoretical determination of the accurate band structure is highly desired.

The normal state of UTe$_2$ can be best explained within a framework of Kondo lattice where the periodic local magnetic moments are screened by spins of conduction electrons~\cite{kondo1964resistance}. 
The Ce-based Kondo lattice has been most widely studied by the dynamical mean field theory combined with density functional theory (DFT+DMFT)~\cite{choi2013observation,jang2020evolution,choi2012temperature,lu2016pressure,kim2019topological,zhu2020kondo}. 
In a recent study by Choi et al.\cite{choi2013observation}, the temperature evolution of the Kondo effect was illustrated by analyzing the spectral function $A(\textrm{\textbf{k}},\omega)$ for CeCoGe$_{2}$. 
The incoherent Ce-4$f$ state is hybridized with conduction electrons ($f$-$c$ hybridization), and the stronger contribution of Ce-4$f$ state near the Fermi level distorts conduction electron bands, resulting in kinks in the bands at the Fermi level. 
%
At low temperatures, the incoherent Ce-4$f$ state forms coherent bands with the renormalized carriers, initiating the coherent Kondo lattice states.
The $f$-$c$ hybridization, which is an indispensable element in Kondo physics, occurs selectively in the type of orbitals. 
In CeCoIn$_{5}$, the selective $f$-$c$ hybridization of two $d$ bands and three crystalline electric-field split $f$ levels gives arise to different dispersion of $d$ bands in the vicinity of the Fermi level~\cite{jang2020evolution}. 
In PuCoGa$_{5}$, the Fermi-liquid behavior appears differently depending on Pu-5$f$ states in $j=$ 5/2 or $j=$ 7/2 multiplet, which are induced by spin-orbit coupling (SOC), resulting in larger Kondo scale for $f$ electrons in $j=$ 5/2 than $j=$ 7/2 multiplet~\cite{brito2018orbital}.

The role of SOC in the Kondo effect is a long standing issue~\cite{zarea2012enhancement}, for which UTe$_2$, comprising heavy elements, is a new avenue.
Obviously, SOC has a significant influence on the band structure in UTe$_2$ as well as its superconductivity. To date, there are two DFT+DMFT studies on UTe$_{2}$ without considering SOC~\cite{xu2019quasi,Miao2020}.
Both show a flat peak hybridized with conduction electrons at 10 K and the peak was suppressed at 200 K, suggesting Kondo coherence at low temperature.


In this work, we ascertain the electronic structure of UTe$_2$ without adjusting parameters and its temperature evolution with/without the inclusion of SOC.
By inclusion of SOC, we found a multi-scaled Kondo effect. 
The SOC causes degenerated U-5$f$ states of the non-SOC system to split into partially occupied states in $j=$ 5/2 and unoccupied states in $j=$ 7/2 multiplet. Within the $j=$ 5/2 subspace, we found two groups of U-5$f$ states, which selectively hybridize with conduction elections $p$ or $d$, giving rise to the orbital-dependent Kondo effect.







{\it Methods -}
We use $ab$-$initio$ linearized quasiparticle self-consistent GW (LQSGW) and dynamical mean field theory (DMFT) method~\cite{tomczak2015qsgw+,choi2016first,choi2019comdmft} to calculated the electronic structure of UTe$_{2}$ which crystallizes into orthorhombic space group Immm (No. 71)~\cite{stowe1996contributions,ikeda2006single}.  
The LQSGW+DMFT is designed as a simplified version of the full GW+EDMFT approach~\cite{sun2002extended,biermann2003first,nilsson2017multitier}. 
It calculates electronic structure by using LQSGW approaches~\cite{kutepov2012electronic,kutepov2017linearized}. 
Then, it corrects the local part of GW self-energy within DMFT~\cite{georges1996dynamical,metzner1989correlated,georges1992hubbard}. 
Within the methodology, the only parameters we adopted from experiments are lattice constants ($a=$ 4.1611, $b=$ 6.1222, $c=$ 13.955 $\textrm{\AA}$)~\cite{ikeda2006single}, and we explicitly calculate all other quantities such as double-counting energy and Coulomb interaction tensor. Then, local self-energies for U-5$f$ and U-6$d$ are obtained by solving two different single impurity models. 
For the LQSGW+DMFT scheme, the code ComDMFT~\cite{choi2019comdmft} was used. For the LQSGW part of the LQSGW+DMFT scheme, the code FlapwMBPT~\cite{kutepov2017linearized} was used.
For the details, please see the supplemental materials.

\begin{table*}[ht]
\caption{Calculated  electron occupation of U-5$f$ orbitals in UTe$_{2}$ at $T=300$ K. U-5$f$ orbitals are labelled for convenience in this work.}\label{table_occ}
\begin{ruledtabular}
\begin{tabular}{c|cccccc|cccccccc}
  $j$ & \multicolumn{6}{c|}{5/2} & \multicolumn{8}{c}{7/2} \\
  \hline
  $j_{z}$ & -2.5 & -1.5 & -0.5 & 0.5 & 1.5 & 2.5 & -3.5 & -2.5 & -1.5 & -0.5 & 0.5 & 1.5 & 2.5 & 3.5 \\
  \hline
  occupation & 0.39 & 0.22 & 0.39 & 0.40 & 0.22 & 0.42 & 0.02 & 0.03 & 0.03 & 0.03 & 0.03 & 0.03 & 0.03 & 0.02 \\ 
  \hline
  label & $f_{1}$ & $f_{2}$ & $f_{3}$ & $f_{4}$& $f_{5}$ & $f_{6}$ & $f_{7}$ & $f_{8}$ & $f_{9}$ & $f_{10}$ & $f_{11}$ & $f_{12}$ & $f_{13}$ & $f_{14}$ \\
\end{tabular}
\end{ruledtabular}
\end{table*}

\begin{figure}[t]
\centering
\includegraphics[width=0.48 \textwidth]{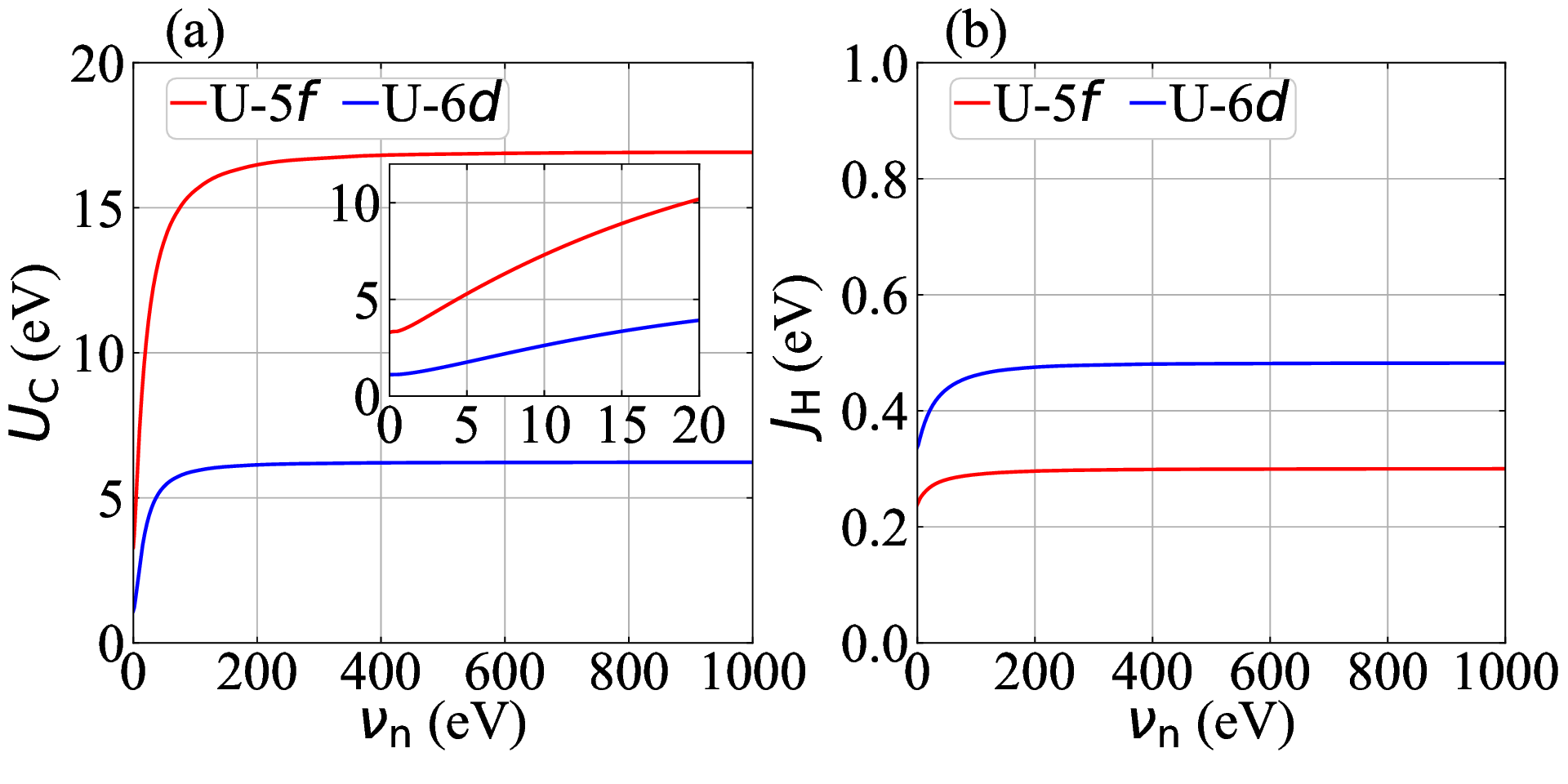}
\caption{\label{Fig_UJ} Calculated (a) $U_{\textrm{C}}$ for U-5$f$ and U-6$d$, and (b) $J_{\textrm{H}}$ for U-5$f$ and U-6$d$ in UTe$_{2}$ with inclusion of SOC. Insets in (a) shows magnified views of $U_{\textrm{C}}$ in low frequency range.}
\end{figure}

{\it Coulomb interaction tensor and electron occupation -}
First, we calculate onsite Coulomb interaction $U_\textrm{C}$ and exchange interaction $J_\textrm{H}$ within the constrained random phase approximation (cRPA)~\cite{aryasetiawan2004frequency,SM}. 
The calculated $U_\textrm{C}$ and $J_\textrm{H}$ for both U-5$f$ and U-6$d$ orbitals with inclusion of SOC are shown in Fig.~\ref{Fig_UJ} (see Fig. S1 for non-SOC results). Both $U_\textrm{C}$ and $J_\textrm{H}$ increase and saturate to the bare unscreened value at high frequency. 

\begin{figure*}[ht]
\centering
\includegraphics[width=0.98 \textwidth]{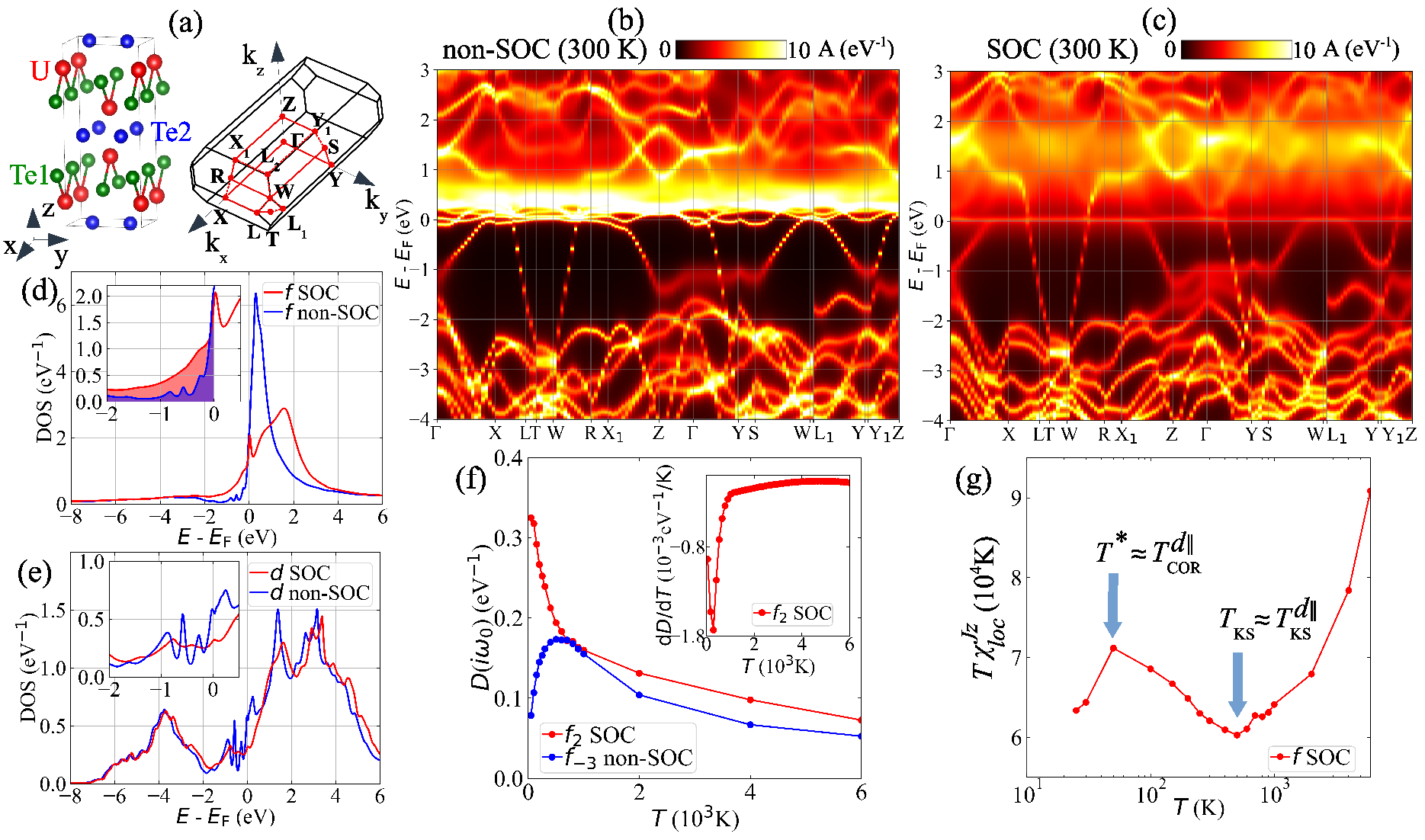}
\caption{\label{Fig_band} (a) Crystal structure and first Brillouin zone of UTe$_{2}$. (b) and (c) Calculated spectral function without and with inclusion of SOC at 300 K. (d) and (e) U-5$f$ and U-6$d$ projected DOSs with and without inclusion of SOC at 300 K. Insets show magnified views near the Fermi level. (f) $f_{2}$ and $f_{-3}$ projected DOSs at the Fermi level for SOC and non-SOC, respectively. Inset shows temperature derivation of $f_{2}$ DOS. (g) $T\chi_{loc}^{J_Z}$ for U-5$f$ with inclusion of SOC. The blue arrows point local minimum and maxima at which scales of Kondo effect were evaluated.}
\end{figure*}

Using the Coulomb interaction tensor, the electron occupancy of U-5$f$ orbitals at 300 K are subsequently calculated including SOC, which is presented in Table~\ref{table_occ}.
The total occupation in U-5$f$ orbitals is 2.27. 
The SOC split U-5$f$ into $j=$ 5/2 and $j=$ 7/2 multiplet, resulting in pushing U-5$f$ states of non-SOC system, which are centered around 0.3 eV above the Fermi level (see Fig.~\ref{Fig_band} (d)).
This interaction gives rise to partially occupied $j=$ 5/2 states and unoccupied $j=$ 7/2 states.
On the contrary, the U atom is strongly oxidized in the non-SOC simulation where the corresponding electron is transferred to Te-5$p$ orbitals, and the resultant occupancy is 1.17 which is significantly smaller than that from the SOC simulation. 
The occupation numbers exhibit no sizable temperature variation from 25 K to 2000 K.
The calculated U-5$f$ occupation with SOC is in good agreement with the measured 5$f^{2}$ configuration~\cite{stowe1997uncommon}.

\begin{figure*}[ht]
\centering
\includegraphics[width=0.98 \textwidth]{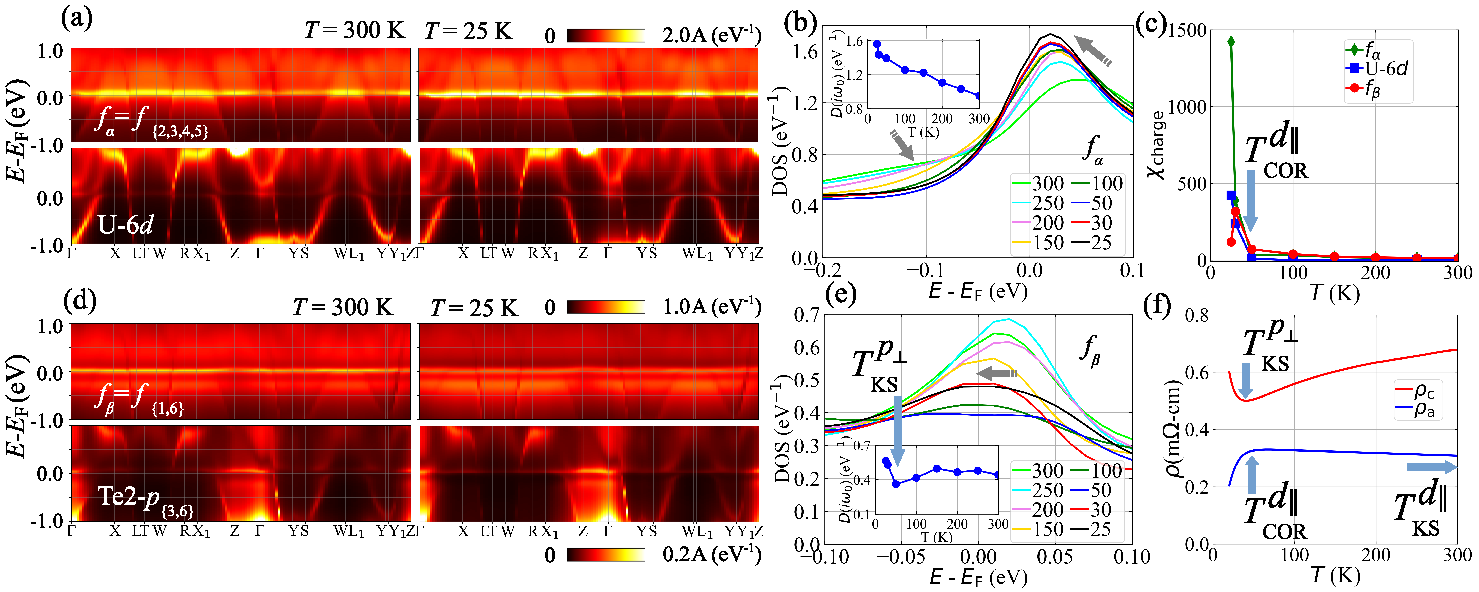}
\caption{\label{Fig_rho}
(a) $f_\alpha$ and U-6$d$ projected spectral functions at 300 (25) K in left (right) panel. (b) $f_\alpha$ projected DOS. The gray arrows denote peaks evolution at temperature lowering. Inset shows $f_\alpha$ DOS at the Fermi level. (c) Charge susceptibility of $f_\alpha$, U-6$d$, and $f_\beta$ orbitals. (d) $f_\beta$ and Te2-$p_{\{3,6\}}$ projected spectral functions at 300 (25) K in left (right) panel. (e) $f_\beta$ projected DOS. The gray arrow denotes the peak center is shifted below the Fermi level. Inset shows DOS at the Fermi level. (f) Corresponding scales of Kondo scattering are denoted on measured resistivity along the a and c axes \cite{Eo2021}. Note that the behavior of resistivity along the b axis (not shown here) is similar with $\rho_\textrm{a}$. In (b) and (e), the temperature unit is K.}
\end{figure*}

{\it Kondo Scattering driven by SOC -} 
To learn the Kondo effect due to the localized U-5$f$ in UTe$_2$, we calculated the spectral function at various temperatures along the high symmetry orientations depicted in Fig.~\ref{Fig_band} (a).
The calculated spectral functions with and without SOC at 300 K are respectively shown in Fig.~\ref{Fig_band} (b) and (c).
Within SOC simulation, the flat heavy electron and hole bands are formed in the vicinity of the Fermi level along every high symmetry line. (see also Fig.~\ref{Fig_rho} (a)). 
Within the non-SOC simulation, the coherent bands near the Fermi energy are comprised of U-5$f$, U-6$d$, and Te-5$p$ orbitals, and the flat bands do not form down to 100 K. 


Figures~\ref{Fig_band} (d) and (e) show the projected DOSs for U-5$f$ and U-6$d$ orbitals, respectively.
The results with and without SOC are shown in red and blue lines, respectively.
In the SOC system, a sharp U-5$f$ peak emerges about 0.02 eV above the Fermi level,
whereas a sharp U-5$f$ peak is centered around 0.3 eV above the Fermi level in the non-SOC system. 
In comparison to U-5$f$ occupation of 1.17 in the non-SOC system, the SOC system manifests a higher U-5$f$ occupation of 2.27 in the larger U-5$f$ DOS below the Fermi level (see inset of Fig.~\ref{Fig_band}(d)). 
In the non-SOC system, three coherent peaks appear between -1 eV and 0 eV, which are hybridized with U-6$d$ as shown in Fig.~\ref{Fig_band} (e).
In contrast, the U-5$f$ and U-6$d$ DOSs below the Fermi level are broad in the SOC system, and the shape of the U-5$f$ spectral function is consistent with incoherent states. 

Figure~\ref{Fig_band} (f) shows the temperature dependence of U-5$f$ projected DOS at the Fermi level, which is determined by
\begin{equation}
  \begin{split}
D(i\omega_{0})=-\frac{1}{\pi}\textrm{Im}G(i\omega_{0})
  \end{split}
\end{equation}
where $\omega_{0}$ is the first Matsubara frequency and $G$ is the calculated local Green's function. 
$D(i\omega_{0})$ for both SOC $f_{2}$ and non-SOC $f_{-3}$ ($l= 3$, $m= -3$) gradually increase upon cooling down to 700 K because of formation of quasiparticle peak where the spectral weight at the Fermi level is transferred from the upper and lower Hubbard bands~\cite{deng2019signatures,choi2013observation}. 
However, the SOC $f_{2}$ and non-SOC $f_{-3}$ exhibit qualitatively different behavior below 700 K.
Whereas $D(i\omega_{0})$ of $f_{-3}$ in the non-SOC system is rapidly reduced, that of $f_{2}$ in the SOC system soars.
$\{$$f_{1}$,$f_{3}$,$f_{4}$,$f_{5}$,$f_{6}$$\}$ in $j=$ 5/2 multiplet show the same temperature evolution with $f_{2}$ down to 300 K. 
We attribute the abruptly enhanced $D(i\omega_{0})$ to the emergence of the Kondo scattering in the SOC system where abundant $f$ DOS near the Fermi level facilitates the $f$-$c$ hybridization. 
The six partially occupied U-5$f$ orbitals in $j=$ 5/2 multiplet are characterized with $-1/2$ spin by Hund's rule, and the total spin moment of the impurity increases, resulting in a relatively higher onset temperature of the Kondo screening process~\cite{koller2005singular}.

To investigate the relevant energy scales of Kondo physics, we calculate the local total angular momentum susceptibility given as
\begin{equation}
  \begin{split}
\chi_{loc}^{J_Z}=\int_0^\beta d\tau\langle J_z(\tau)J_z(0)\rangle.
  \end{split}
\end{equation}
Figure~\ref{Fig_band} (g) shows the product of temperature and $\chi_{loc}^{J_Z}$ of U-5$f$ as a function of temperature.
The plot reveals two characteristic temperatures of Kondo lattice in UTe$_2$: Kondo scattering temperatures ($T_{\textrm{KS}}$) and lattice coherence temperature ($T^*$). 
The high temperature $T_{\textrm{KS}}$ below which $\chi_{loc}^{J_Z}$ deviates from the Curie-Weiss behaviors indicates the onset of the Kondo scattering process. 
The low temperature $T^*$ with a local maximum of $\chi_{loc}^{J_Z}$ suggests the crossover of incoherent-to-coherent scattering. 
Our estimation of $T^*$ $\simeq 50$ K and $T_{\textrm{KS}}$ $>$ 300 K is consistent with the negative slope of $ab$-plane resistivity between 50 K and 300 K~\cite{Eo2021}.
$T_{\textrm{KS}}$ in UTe$_2$ is much higher than $\sim$ 200 K in CeCoIn$_{5}$~\cite{jang2020evolution} and CeCoGe$_{2}$~\cite{choi2013observation}, and we attribute the higher $T_{\textrm{KS}}$ to the larger total $f$ valence electrons in U-5$f$ than Ce-4$f$ and strong SOC in UTe$_2$.


{\it Orbital selective Kondo Scattering -}
To elucidate the transport properties in UTe$_{2}$ between $T=$ 25 K  and 300 K, we analyze the SOC system in depth. 
The crystal electric field and SOC lift the degeneracy of the U-5$f$ orbitals, grossly forming two groups, $f_\alpha$=$\{$$f_{2}$,$f_{3}$,$f_{4}$,$f_{5}$$\}$ and $f_\beta$=$\{$$f_{1}$,$f_{6}$$\}$ (see Table~\ref{table_occ}), with an energy separation of $\sim$0.12 eV.
These two groups show distinct electronic structures in this temperature range in terms of DOS, spectral functions, and charge susceptibility.
We found that $f_\alpha$ and $f_\beta$ are hybridized with U-6$d$ and Te2-$p_{\{{3,6\}}}$ states, respectively.
Here, Te2-$p_{\{{3,6\}}}$ includes Te2-5$p$ states of ($j=$ 3/2, $j_{z}=$ -3/2) and ($j=$ 3/2, $j_{z}=$ 3/2).

We computed orbital-resolved spectral functions, and the hybridization between $f_\alpha$ and U-6$d$ is presented in Fig.~\ref{Fig_rho} (a). 
The kink-like structure appears at the intersection of the two bands, which is consistent with the Kondo effect~\cite{choi2013observation}.
The abrupt change in the U-6$d$ conduction bands dispersion at the Fermi level is clearly visible along the $\Gamma$-X, X$_{1}$-Z, S-W, and L$_{1}$-Y symmetry lines, all of which belong to the $ab$-plane.
Thus, these hybridizations will affect electrical transport in the $ab$-plane. 
We define the onset temperature of Kondo scattering due to hybridization between $f_\alpha$ and U-6$d$, $T_{\textrm{KS}}^{d_{\parallel}}\approx 500$ K at which the kink-like structure starts appearing at the Fermi level. Note that $T_{\textrm{KS}}^{d_{\parallel}}$ $\approx$ $T_\textrm{KS}$, indicating $f_\alpha$-orbitals are responsible for the high temperature Kondo effect.
Below $T$ = 400 K, the heavy mass dispersion appears, and the first excited state is disconnected, indicating the progression of the active Kondo scattering.
At $T$ = 25 K, the incoherent $f_\alpha$-bands below the Fermi level are absent, and the coherent $f_\alpha$ peak above the Fermi level is enhanced near the Fermi level, resulting in a single coherent peak (see Fig.~\ref{Fig_rho} (a), (b), and Fig. S2). 
The formation of coherent $f$ bands near the Fermi level signals renormalization of the carriers~\cite{choi2013observation} and indicates the coherent Kondo lattice in UTe$_{2}$~\cite{jang2020evolution}. 
We determine the crossover temperature $T_\textrm{cor}^{d_{\parallel}}$ (often denoted by $T^*$) from the incoherent scattering to the coherent Kondo lattice by employing the charge susceptibility,
\begin{equation}
  \begin{split}
\chi_\textrm{charge}=\sum_{i}\int_0^\beta d\tau \langle n_{i}(\tau) n_{i}(0)\rangle - \beta \langle  n_{i} \rangle^{2}
  \end{split}
\end{equation}
where $n_{i}$ is the occupation of $i$-th orbital, which evaluates charge fluctuation~\cite{koga2005spin}. 
Figure~\ref{Fig_rho} (c) shows the temperature dependent $\chi_\textrm{charge}$ of $f_\alpha$ (green) and U-6$d$ (blue) which abruptly surge below $T_\textrm{cor}^{d_{\parallel}}=50$ K simultaneously. 
Our estimated $T_{\textrm{KS}}^{d_{\parallel}}=$ 500 K and $T_\textrm{cor}^{d_{\parallel}}=$ 50 K for the $f_\alpha$ and U-6$d$ hybridization are in good agreement with experimental resistivity along the $a$-and $b$-axis, where $\rho_{a}$ and $\rho_{b}$ show $d\rho/dT<0$ between 50 K and 300 K and start to drop rapidly around 50 K~\cite{Eo2021} as shown in Fig.~\ref{Fig_rho} (f).
Te1-5$p$ orbitals, hybridized with U-5$f$ and U-6$d$, constitute the conduction bands along the X-L, W-R, $\Gamma$-Y, and Y$_{1}$-Z symmetry lines. However, we did not find any evidence of the Kondo effect, and the hybridized bands remain metallic at high temperatures.

Below we focus on $f_\beta$-orbitals.
Figure~\ref{Fig_rho} (d) shows the $f_{\beta}$ orbital-resolved spectral functions where $f_{\beta}$ is hybridized with Te2-$p_{\{{3,6\}}}$ along Z-$\Gamma$ in the vicinity of the Fermi level (see also Fig. S3 (a)).
This result indicates that the conduction electron along the $c$-axis is originated from Te2-$p_{\{{3,6\}}}$.
The DOS of Te2-$p_{\{{3,6\}}}$ (data not shown) consists of spectral weight at the Fermi level over the entire temperature range and shows a small but notable increase as lowering the temperature. 
Figure~\ref{Fig_rho} (e) shows $f_\beta$-projected DOS where a single coherent peak is present slightly above the Fermi level at high temperatures.
The peak position approaches the Fermi level upon cooling, accompanied by a gradual reduction of the peak height.
At $T=$ 50 K, another peak appears below the Fermi level. 
The two peaks merge at $T<50$ K, forming a single, enhanced peak at the Fermi level  (see also Fig. S2).
The temperature evolution of DOS of $f_\beta$ at the Fermi level can shed light on the transport properties, and the inset of Fig.~\ref{Fig_rho} (e) shows temperature evolution of $D(i\omega_{0})$ of $f_\beta$.
Whereas $D(i\omega_{0})$ of $f_\alpha$ monotonically increases upon cooling (see inset of Fig.~\ref{Fig_rho} (b)), $D(i\omega_{0})$ of $f_{\beta}$ is nearly temperature-independent between 150 K and 300 K.
It exhibits a local minimum at 50 K before sharply increasing at lower temperatures. 
We attribute our observations to the Kondo scattering involving $f_{\beta}$ and Te2-$p_{\{{3,6\}}}$ at $T=$ 50 K (see Fig. S3 (b)). 
We thus define $T_{\textrm{KS}}^{p_{\perp}}=$ 50 K as the onset temperature of the Kondo effect along the $c$-aixs, and, $\chi_\textrm{charge}$ of $f_{\beta}$ manifests a sudden drop below $T=$ 30 K as shown in Fig.~\ref{Fig_rho} (c). 
The hybridization between $f_{\beta}$ and Te2-$p_{\{{3,6\}}}$ is responsible for the $c$-axis transport, and our result is consistent with the recent experiment where $\rho_{c}$ is metallic in $\sim$ 50 K $<$ $T$ $<$ 300 K and exhibits an upturn below $\sim$ 50 K~\cite{Eo2021} as shown in Fig.~\ref{Fig_rho} (f). 

\begin{figure}[ht]
\centering
\includegraphics[width=0.50 \textwidth]{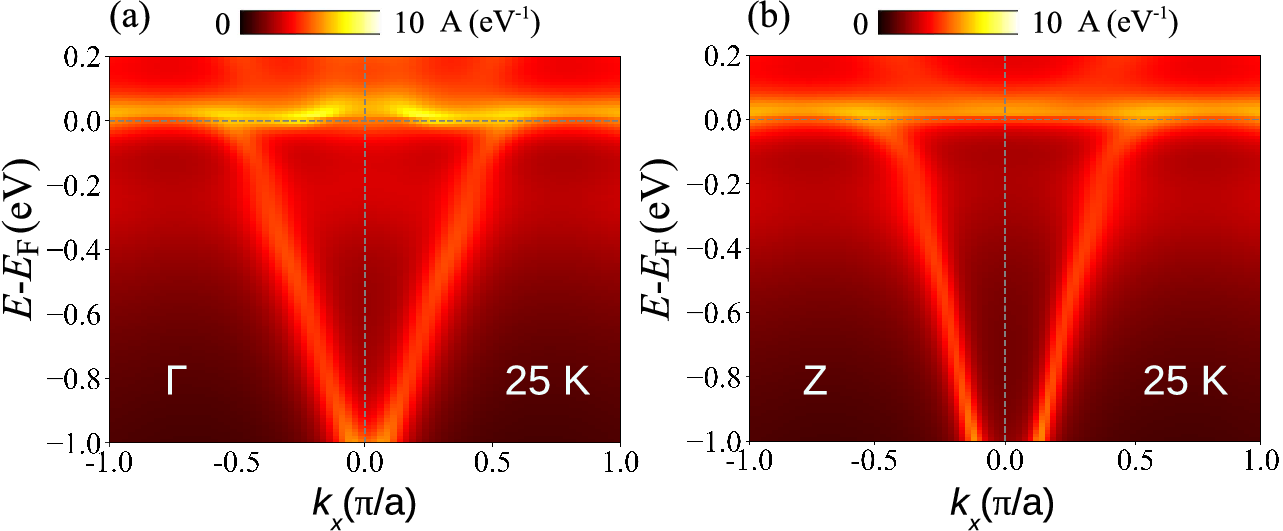}
\caption{\label{Fig_arpes} Momentum-resolved total spectral function along the (a) X-$\Gamma$-X and (b) X$_\textrm{1}$-Z-X$_\textrm{1}$. The simulation temperature is 25 K for both.}
\end{figure}

{\it ARPES-}
We compare our calculated spectral function at $T$ $=$ 25 K to recent ARPES measurements at $T$ $=$ 20 K~\cite{Fujimori2019,Miao2020}. The calculated spectral functions along the $\Gamma$-X and Z-$\textrm{X}_1$ consist of one parabolic band and incoherent spectral weights, which agree with both measured ARPES spectra in the binding energy of $E_\textrm{B} <$ 1.0 eV along the same symmetry lines. Particularly, dispersive Kondo resonance peaks near the Fermi level (which is not presented in the non-SOC spectral function.) appear in the ARPES~\cite{Fujimori2019}. There are incoherent $f_\beta$ bands in $-$ 0.2 eV $>$ $E-E_\textrm{F}$ $>$ $-$ 0.4 eV. The corresponding bands appear in  $-$ 0.4 eV $>$ $E-E_\textrm{F}$ $>$ $-$ 0.6 eV in both ARPES data. As shown in Fig.~\ref{Fig_rho} (d), the spectral weight is more pronounced at lower energy down to 25 K from 300 K. Also, transport measurement shows abrupt change on the resistivity below 50 K (see Fig.~\ref{Fig_rho} (f)), indicating the rapid change on the electronic structure. Thus, we suggest that the variance may partly be caused by the temperature difference between simulation and ARPES measurement.   

{\it Discussion-}
\begin{figure}[ht]
\centering
\includegraphics[width=0.50 \textwidth]{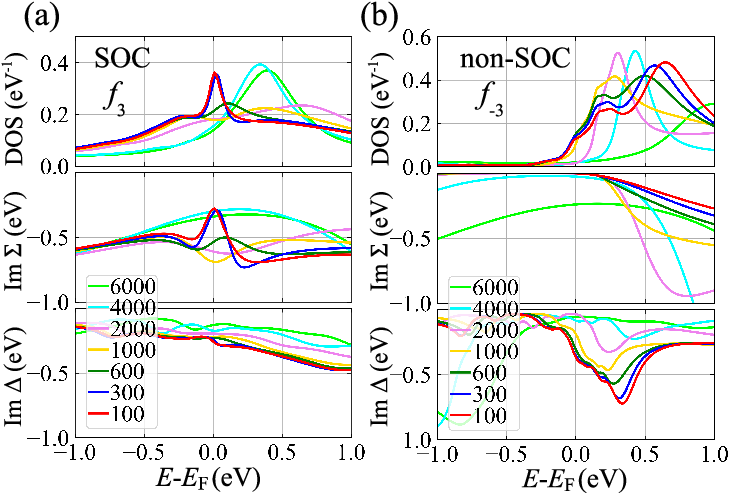}
\caption{\label{Fig_sig} Calculated density of states, the imaginary part of the self-energy and hybridization function of (a) $f_{3}$ in SOC and (b) $f_{-3}$ in non-SOC system. The simulation temperature unit is K. The zero of energy is set at the Fermi level.}
\end{figure}
We have shown distinct U-5$f$ peaks in the vicinity of the Fermi level for both SOC and non-SOC systems at 300 K in Fig.~\ref{Fig_band} (d). 
The origin of characteristic $f$ peak features and its subsequent temperature evolution can be identified with the self-energy and hybridization function.
In Fig.~\ref{Fig_sig} (a) for SOC system, the $f_{3}$ self-energies exhibit strong poles at the Fermi level below 300 K due to the Kondo screening process. While this gives rise to the prominent $f$ peak at the Fermi level, the hybridization functions did not show notable changes at the Fermi level with respect to temperatures. 
In contrary, as shown in Fig.~\ref{Fig_sig} (b) for non-SOC system, self-energies do not have a pole at the Fermi level. Instead, the broader $f$ bands in the vicinity of the Fermi level are developed by orbital hybridization as indicated by pronounced hybridization functions. The diverging hybridization functions as lowering temperature indicates enhanced electron or hole hopping between $f$ and adjacent orbitals.

\begin{figure}[ht]
\centering
\includegraphics[width=0.50 \textwidth]{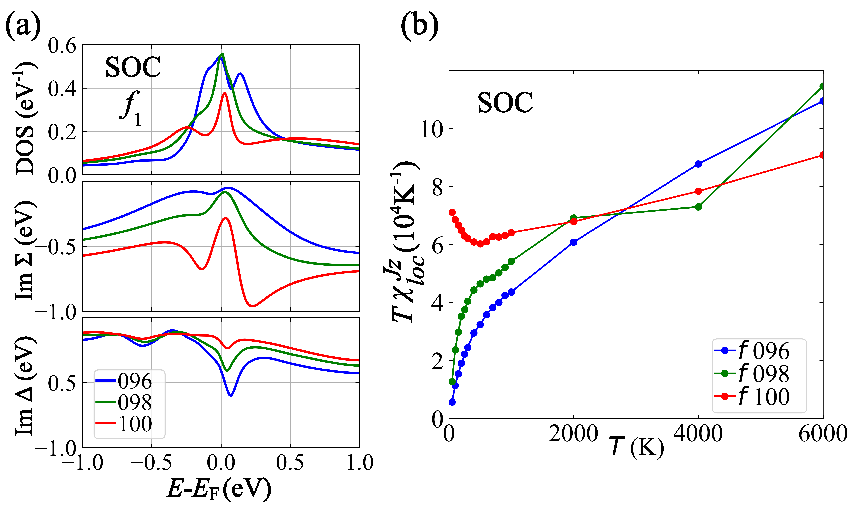}
\caption{\label{Fig_pressure} (a) Calculated density of states, the imaginary part of the self-energy and hybridization function of $f_{1}$ in SOC system at 300 K. The structures are denoted by 100, 098, and 096 for 0, 2 and 4 percent reduced experimental lattice constants, respectively. (b) $T\chi_{loc}^{J_Z}$ for U-5$f$ with inclusion of SOC.}
\end{figure}
The understanding of the electronic structure at high pressure may provide important clues about the origin of enhanced $T_c$ at high pressure\cite{Ran2020,Knafo2021}.
We suggest that U-5$f$ electron is less localized at high pressure, resulting in a suppressed Kondo effect which is reminiscent of the non-SOC system.
In this work, UTe$_2$ at high pressure is studied with simulations, where structures were constructed by adopting reduced experimental lattice constants without geometry relaxation. The geometry optimization using many-body methods (with proper electronic structure) is not yet feasible. Our simulations do not show quantitative result, but reveal distinct features due to reduced atomic distance. 
The modeled structures are denoted by 100, 098, and 096 for 0, 2 and 4 percent reduced experimental lattice constants, respectively.
As shown in Fig.~\ref{Fig_pressure} (a), pole strengths in the self-energy decreased in association with a reduced lattice constants, unlike enhanced divergence of the hybridization functions.
As shown in Fig.~\ref{Fig_pressure} (b), we do not find a local minimum in  $\chi_{loc}^{J_Z}$, which indicates the onset of the Kondo scattering process, down to 50 K for 098 and 096 simulations.
Reducing lattice constants causes more overlap between $f$ and adjacent orbitals. It enhances $f$ electron hopping between the orbitals, resulting in decreased local magnetic moments.
Our results imply the Kondo effect is weakened at high pressure. Therefore, it is predicted that the onset of resistivity upturn by the Kondo effect may shift down to the lower temperature with respect to increasing pressure. 

{\it Summary -} 
We performed first-principles simulations to investigate the electronic structure of heavy fermion superconductor UTe$_{2}$. We found that the inclusion of SOC is necessary to reproduce the observed occupancy of the 5$f$ orbital-configuration, anomalous temperature-dependence of electrical resistivity by Kondo effect, and band structure near the Fermi level measured by ARPES.
In regard to electron correlations in heavy fermion compounds, we note a number of DFT+DMFT studies for Ce-and Pu-based compounds~\cite{choi2013observation,jang2020evolution,choi2012temperature,brito2018orbital,lu2021temperature}, where conduction electron  ($d$ or $p$) dependent correlations is not found. As such, our discovery suggests that UTe$_{2}$ has complex electronic structure, which manifests unprecedented orbital-selective hybridization among $f_\textrm{\{1,6\}}$, $f_\textrm{\{2,3,4,5\}}$, $d$, and $p$ orbitals. 
Our description of rich physics of the normal state in UTe$_{2}$ may provide key knowledge for the understanding of unconventional superconductivity.

{\it Acknowledgements -} 
The authors would like to thank Yun Suk Eo and Johnpierre Paglione for sharing the electrical transport data and insightful discussion. We acknowledge the High Performance Computing Center (HPCC) at Texas Tech University for providing computational resources that have contributed to the research results reported within this paper. S.C. was supported by the U.S. Department of Energy, Office of Science, Basic Energy Sciences as a part of the Computational Materials Science Program. Coulomb interaction tensor calculation used resources of the National Energy Research Scientific Computing Center (NERSC), a U.S. Department of Energy Office of Science User Facility operated under Contract No. DE-AC02- 05CH11231 

\bibliography{ref}

\begin{thebibliography}{46}%
\makeatletter
\providecommand \@ifxundefined [1]{%
 \@ifx{#1\undefined}
}%
\providecommand \@ifnum [1]{%
 \ifnum #1\expandafter \@firstoftwo
 \else \expandafter \@secondoftwo
 \fi
}%
\providecommand \@ifx [1]{%
 \ifx #1\expandafter \@firstoftwo
 \else \expandafter \@secondoftwo
 \fi
}%
\providecommand \natexlab [1]{#1}%
\providecommand \enquote  [1]{``#1''}%
\providecommand \bibnamefont  [1]{#1}%
\providecommand \bibfnamefont [1]{#1}%
\providecommand \citenamefont [1]{#1}%
\providecommand \href@noop [0]{\@secondoftwo}%
\providecommand \href [0]{\begingroup \@sanitize@url \@href}%
\providecommand \@href[1]{\@@startlink{#1}\@@href}%
\providecommand \@@href[1]{\endgroup#1\@@endlink}%
\providecommand \@sanitize@url [0]{\catcode `\\12\catcode `\$12\catcode
  `\&12\catcode `\#12\catcode `\^12\catcode `\_12\catcode `\%12\relax}%
\providecommand \@@startlink[1]{}%
\providecommand \@@endlink[0]{}%
\providecommand \url  [0]{\begingroup\@sanitize@url \@url }%
\providecommand \@url [1]{\endgroup\@href {#1}{\urlprefix }}%
\providecommand \urlprefix  [0]{URL }%
\providecommand \Eprint [0]{\href }%
\providecommand \doibase [0]{https://doi.org/}%
\providecommand \selectlanguage [0]{\@gobble}%
\providecommand \bibinfo  [0]{\@secondoftwo}%
\providecommand \bibfield  [0]{\@secondoftwo}%
\providecommand \translation [1]{[#1]}%
\providecommand \BibitemOpen [0]{}%
\providecommand \bibitemStop [0]{}%
\providecommand \bibitemNoStop [0]{.\EOS\space}%
\providecommand \EOS [0]{\spacefactor3000\relax}%
\providecommand \BibitemShut  [1]{\csname bibitem#1\endcsname}%
\let\auto@bib@innerbib\@empty
\bibitem [{\citenamefont {Aoki}\ \emph
  {et~al.}(2019{\natexlab{a}})\citenamefont {Aoki}, \citenamefont {Ishida},\
  and\ \citenamefont {Flouquet}}]{Aoki2019review}%
  \BibitemOpen
  \bibfield  {author} {\bibinfo {author} {\bibfnamefont {D.}~\bibnamefont
  {Aoki}}, \bibinfo {author} {\bibfnamefont {K.}~\bibnamefont {Ishida}},\ and\
  \bibinfo {author} {\bibfnamefont {J.}~\bibnamefont {Flouquet}},\ }\href
  {https://doi.org/10.7566/JPSJ.88.022001} {\bibfield  {journal} {\bibinfo
  {journal} {Journal of the Physical Society of Japan}\ }\textbf {\bibinfo
  {volume} {88}},\ \bibinfo {pages} {022001} (\bibinfo {year}
  {2019}{\natexlab{a}})}\BibitemShut {NoStop}%
\bibitem [{\citenamefont {Aoki}\ \emph {et~al.}(2001)\citenamefont {Aoki},
  \citenamefont {Huxley}, \citenamefont {Ressouche}, \citenamefont
  {Braithwaite}, \citenamefont {Flouquet}, \citenamefont {Brison},
  \citenamefont {Lhotel},\ and\ \citenamefont {Paulsen}}]{Aoki2001}%
  \BibitemOpen
  \bibfield  {author} {\bibinfo {author} {\bibfnamefont {D.}~\bibnamefont
  {Aoki}}, \bibinfo {author} {\bibfnamefont {A.}~\bibnamefont {Huxley}},
  \bibinfo {author} {\bibfnamefont {E.}~\bibnamefont {Ressouche}}, \bibinfo
  {author} {\bibfnamefont {D.}~\bibnamefont {Braithwaite}}, \bibinfo {author}
  {\bibfnamefont {J.}~\bibnamefont {Flouquet}}, \bibinfo {author}
  {\bibfnamefont {J.-P.}\ \bibnamefont {Brison}}, \bibinfo {author}
  {\bibfnamefont {E.}~\bibnamefont {Lhotel}},\ and\ \bibinfo {author}
  {\bibfnamefont {C.}~\bibnamefont {Paulsen}},\ }\href
  {https://doi.org/10.1038/35098048} {\bibfield  {journal} {\bibinfo  {journal}
  {Nature}\ }\textbf {\bibinfo {volume} {413}},\ \bibinfo {pages} {613}
  (\bibinfo {year} {2001})}\BibitemShut {NoStop}%
\bibitem [{\citenamefont {Huy}\ \emph {et~al.}(2007)\citenamefont {Huy},
  \citenamefont {Gasparini}, \citenamefont {de~Nijs}, \citenamefont {Huang},
  \citenamefont {Klaasse}, \citenamefont {Gortenmulder}, \citenamefont
  {de~Visser}, \citenamefont {Hamann}, \citenamefont {G\"orlach},\ and\
  \citenamefont {L\"ohneysen}}]{Huy2007}%
  \BibitemOpen
  \bibfield  {author} {\bibinfo {author} {\bibfnamefont {N.~T.}\ \bibnamefont
  {Huy}}, \bibinfo {author} {\bibfnamefont {A.}~\bibnamefont {Gasparini}},
  \bibinfo {author} {\bibfnamefont {D.~E.}\ \bibnamefont {de~Nijs}}, \bibinfo
  {author} {\bibfnamefont {Y.}~\bibnamefont {Huang}}, \bibinfo {author}
  {\bibfnamefont {J.~C.~P.}\ \bibnamefont {Klaasse}}, \bibinfo {author}
  {\bibfnamefont {T.}~\bibnamefont {Gortenmulder}}, \bibinfo {author}
  {\bibfnamefont {A.}~\bibnamefont {de~Visser}}, \bibinfo {author}
  {\bibfnamefont {A.}~\bibnamefont {Hamann}}, \bibinfo {author} {\bibfnamefont
  {T.}~\bibnamefont {G\"orlach}},\ and\ \bibinfo {author} {\bibfnamefont
  {H.~v.}\ \bibnamefont {L\"ohneysen}},\ }\href
  {https://doi.org/10.1103/PhysRevLett.99.067006} {\bibfield  {journal}
  {\bibinfo  {journal} {Phys. Rev. Lett.}\ }\textbf {\bibinfo {volume} {99}},\
  \bibinfo {pages} {067006} (\bibinfo {year} {2007})}\BibitemShut {NoStop}%
\bibitem [{\citenamefont {Ran}\ \emph {et~al.}(2019)\citenamefont {Ran},
  \citenamefont {Eckberg}, \citenamefont {Ding}, \citenamefont {Furukawa},
  \citenamefont {Metz}, \citenamefont {Saha}, \citenamefont {Liu},
  \citenamefont {Zic}, \citenamefont {Kim}, \citenamefont {Paglione},\ and\
  \citenamefont {Butch}}]{Ran2019science}%
  \BibitemOpen
  \bibfield  {author} {\bibinfo {author} {\bibfnamefont {S.}~\bibnamefont
  {Ran}}, \bibinfo {author} {\bibfnamefont {C.}~\bibnamefont {Eckberg}},
  \bibinfo {author} {\bibfnamefont {Q.-P.}\ \bibnamefont {Ding}}, \bibinfo
  {author} {\bibfnamefont {Y.}~\bibnamefont {Furukawa}}, \bibinfo {author}
  {\bibfnamefont {T.}~\bibnamefont {Metz}}, \bibinfo {author} {\bibfnamefont
  {S.~R.}\ \bibnamefont {Saha}}, \bibinfo {author} {\bibfnamefont {I.-L.}\
  \bibnamefont {Liu}}, \bibinfo {author} {\bibfnamefont {M.}~\bibnamefont
  {Zic}}, \bibinfo {author} {\bibfnamefont {H.}~\bibnamefont {Kim}}, \bibinfo
  {author} {\bibfnamefont {J.}~\bibnamefont {Paglione}},\ and\ \bibinfo
  {author} {\bibfnamefont {N.~P.}\ \bibnamefont {Butch}},\ }\href
  {https://doi.org/10.1126/science.aav8645} {\bibfield  {journal} {\bibinfo
  {journal} {Science}\ }\textbf {\bibinfo {volume} {365}},\ \bibinfo {pages}
  {684} (\bibinfo {year} {2019})}\BibitemShut {NoStop}%
\bibitem [{\citenamefont {Rosa}\ \emph {et~al.}(2021)\citenamefont {Rosa},
  \citenamefont {Weiland}, \citenamefont {Fender}, \citenamefont {Scott},
  \citenamefont {Ronning}, \citenamefont {Thompson}, \citenamefont {Bauer},\
  and\ \citenamefont {Thomas}}]{Rosa2021}%
  \BibitemOpen
  \bibfield  {author} {\bibinfo {author} {\bibfnamefont {P.~F.~S.}\
  \bibnamefont {Rosa}}, \bibinfo {author} {\bibfnamefont {A.}~\bibnamefont
  {Weiland}}, \bibinfo {author} {\bibfnamefont {S.~S.}\ \bibnamefont {Fender}},
  \bibinfo {author} {\bibfnamefont {B.~L.}\ \bibnamefont {Scott}}, \bibinfo
  {author} {\bibfnamefont {F.}~\bibnamefont {Ronning}}, \bibinfo {author}
  {\bibfnamefont {J.~D.}\ \bibnamefont {Thompson}}, \bibinfo {author}
  {\bibfnamefont {E.~D.}\ \bibnamefont {Bauer}},\ and\ \bibinfo {author}
  {\bibfnamefont {S.~M.}\ \bibnamefont {Thomas}},\ }\href@noop {} {\bibfield
  {journal} {\bibinfo  {journal} {arXiv2110.06200}\ } (\bibinfo {year}
  {2021})}\BibitemShut {NoStop}%
\bibitem [{\citenamefont {Aoki}\ \emph {et~al.}(2021)\citenamefont {Aoki},
  \citenamefont {Brison}, \citenamefont {Flouquet}, \citenamefont {Ishida},
  \citenamefont {Knebel}, \citenamefont {Tokunaga},\ and\ \citenamefont
  {Yanase}}]{Aoki2021review}%
  \BibitemOpen
  \bibfield  {author} {\bibinfo {author} {\bibfnamefont {D.}~\bibnamefont
  {Aoki}}, \bibinfo {author} {\bibfnamefont {J.~P.}\ \bibnamefont {Brison}},
  \bibinfo {author} {\bibfnamefont {J.}~\bibnamefont {Flouquet}}, \bibinfo
  {author} {\bibfnamefont {K.}~\bibnamefont {Ishida}}, \bibinfo {author}
  {\bibfnamefont {G.}~\bibnamefont {Knebel}}, \bibinfo {author} {\bibfnamefont
  {Y.}~\bibnamefont {Tokunaga}},\ and\ \bibinfo {author} {\bibfnamefont
  {Y.}~\bibnamefont {Yanase}},\ }\href@noop {} {\bibfield  {journal} {\bibinfo
  {journal} {arXiv:2110.10451}\ } (\bibinfo {year} {2021})}\BibitemShut
  {NoStop}%
\bibitem [{\citenamefont {Ran}\ \emph {et~al.}(2020)\citenamefont {Ran},
  \citenamefont {Kim}, \citenamefont {Liu}, \citenamefont {Saha}, \citenamefont
  {Hayes}, \citenamefont {Metz}, \citenamefont {Eo}, \citenamefont {Paglione},\
  and\ \citenamefont {Butch}}]{Ran2020}%
  \BibitemOpen
  \bibfield  {author} {\bibinfo {author} {\bibfnamefont {S.}~\bibnamefont
  {Ran}}, \bibinfo {author} {\bibfnamefont {H.}~\bibnamefont {Kim}}, \bibinfo
  {author} {\bibfnamefont {I.-L.}\ \bibnamefont {Liu}}, \bibinfo {author}
  {\bibfnamefont {S.~R.}\ \bibnamefont {Saha}}, \bibinfo {author}
  {\bibfnamefont {I.}~\bibnamefont {Hayes}}, \bibinfo {author} {\bibfnamefont
  {T.}~\bibnamefont {Metz}}, \bibinfo {author} {\bibfnamefont {Y.~S.}\
  \bibnamefont {Eo}}, \bibinfo {author} {\bibfnamefont {J.}~\bibnamefont
  {Paglione}},\ and\ \bibinfo {author} {\bibfnamefont {N.~P.}\ \bibnamefont
  {Butch}},\ }\href {https://doi.org/10.1103/PhysRevB.101.140503} {\bibfield
  {journal} {\bibinfo  {journal} {Phys. Rev. B}\ }\textbf {\bibinfo {volume}
  {101}},\ \bibinfo {pages} {140503} (\bibinfo {year} {2020})}\BibitemShut
  {NoStop}%
\bibitem [{\citenamefont {Knafo}\ \emph {et~al.}(2021)\citenamefont {Knafo},
  \citenamefont {Nardone}, \citenamefont {Vali{\v s}ka}, \citenamefont
  {Zitouni}, \citenamefont {Lapertot}, \citenamefont {Aoki}, \citenamefont
  {Knebel},\ and\ \citenamefont {Braithwaite}}]{Knafo2021}%
  \BibitemOpen
  \bibfield  {author} {\bibinfo {author} {\bibfnamefont {W.}~\bibnamefont
  {Knafo}}, \bibinfo {author} {\bibfnamefont {M.}~\bibnamefont {Nardone}},
  \bibinfo {author} {\bibfnamefont {M.}~\bibnamefont {Vali{\v s}ka}}, \bibinfo
  {author} {\bibfnamefont {A.}~\bibnamefont {Zitouni}}, \bibinfo {author}
  {\bibfnamefont {G.}~\bibnamefont {Lapertot}}, \bibinfo {author}
  {\bibfnamefont {D.}~\bibnamefont {Aoki}}, \bibinfo {author} {\bibfnamefont
  {G.}~\bibnamefont {Knebel}},\ and\ \bibinfo {author} {\bibfnamefont
  {D.}~\bibnamefont {Braithwaite}},\ }\href
  {https://doi.org/10.1038/s42005-021-00545-z} {\bibfield  {journal} {\bibinfo
  {journal} {Communications Physics}\ }\textbf {\bibinfo {volume} {4}},\
  \bibinfo {pages} {40} (\bibinfo {year} {2021})}\BibitemShut {NoStop}%
\bibitem [{\citenamefont {Aoki}\ \emph
  {et~al.}(2019{\natexlab{b}})\citenamefont {Aoki}, \citenamefont {Nakamura},
  \citenamefont {Honda}, \citenamefont {Li}, \citenamefont {Homma},
  \citenamefont {Shimizu}, \citenamefont {Sato}, \citenamefont {Knebel},
  \citenamefont {Brison}, \citenamefont {Pourret} \emph {et~al.}}]{Aoki2019}%
  \BibitemOpen
  \bibfield  {author} {\bibinfo {author} {\bibfnamefont {D.}~\bibnamefont
  {Aoki}}, \bibinfo {author} {\bibfnamefont {A.}~\bibnamefont {Nakamura}},
  \bibinfo {author} {\bibfnamefont {F.}~\bibnamefont {Honda}}, \bibinfo
  {author} {\bibfnamefont {D.}~\bibnamefont {Li}}, \bibinfo {author}
  {\bibfnamefont {Y.}~\bibnamefont {Homma}}, \bibinfo {author} {\bibfnamefont
  {Y.}~\bibnamefont {Shimizu}}, \bibinfo {author} {\bibfnamefont {Y.~J.}\
  \bibnamefont {Sato}}, \bibinfo {author} {\bibfnamefont {G.}~\bibnamefont
  {Knebel}}, \bibinfo {author} {\bibfnamefont {J.-P.}\ \bibnamefont {Brison}},
  \bibinfo {author} {\bibfnamefont {A.}~\bibnamefont {Pourret}}, \emph
  {et~al.},\ }\href@noop {} {\bibfield  {journal} {\bibinfo  {journal} {Journal
  of the Physical Society of Japan}\ }\textbf {\bibinfo {volume} {88}},\
  \bibinfo {pages} {043702} (\bibinfo {year} {2019}{\natexlab{b}})}\BibitemShut
  {NoStop}%
\bibitem [{\citenamefont {Trovarelli}\ \emph {et~al.}(2000)\citenamefont
  {Trovarelli}, \citenamefont {Geibel}, \citenamefont {Mederle}, \citenamefont
  {Langhammer}, \citenamefont {Grosche}, \citenamefont {Gegenwart},
  \citenamefont {Lang}, \citenamefont {Sparn},\ and\ \citenamefont
  {Steglich}}]{Trovarelli2000}%
  \BibitemOpen
  \bibfield  {author} {\bibinfo {author} {\bibfnamefont {O.}~\bibnamefont
  {Trovarelli}}, \bibinfo {author} {\bibfnamefont {C.}~\bibnamefont {Geibel}},
  \bibinfo {author} {\bibfnamefont {S.}~\bibnamefont {Mederle}}, \bibinfo
  {author} {\bibfnamefont {C.}~\bibnamefont {Langhammer}}, \bibinfo {author}
  {\bibfnamefont {F.~M.}\ \bibnamefont {Grosche}}, \bibinfo {author}
  {\bibfnamefont {P.}~\bibnamefont {Gegenwart}}, \bibinfo {author}
  {\bibfnamefont {M.}~\bibnamefont {Lang}}, \bibinfo {author} {\bibfnamefont
  {G.}~\bibnamefont {Sparn}},\ and\ \bibinfo {author} {\bibfnamefont
  {F.}~\bibnamefont {Steglich}},\ }\href
  {https://doi.org/10.1103/PhysRevLett.85.626} {\bibfield  {journal} {\bibinfo
  {journal} {Phys. Rev. Lett.}\ }\textbf {\bibinfo {volume} {85}},\ \bibinfo
  {pages} {626} (\bibinfo {year} {2000})}\BibitemShut {NoStop}%
\bibitem [{\citenamefont {Petrovic}\ \emph {et~al.}(2001)\citenamefont
  {Petrovic}, \citenamefont {Pagliuso}, \citenamefont {Hundley}, \citenamefont
  {Movshovich}, \citenamefont {Sarrao}, \citenamefont {Thompson}, \citenamefont
  {Fisk},\ and\ \citenamefont {Monthoux}}]{Petrovic2001}%
  \BibitemOpen
  \bibfield  {author} {\bibinfo {author} {\bibfnamefont {C.}~\bibnamefont
  {Petrovic}}, \bibinfo {author} {\bibfnamefont {P.~G.}\ \bibnamefont
  {Pagliuso}}, \bibinfo {author} {\bibfnamefont {M.~F.}\ \bibnamefont
  {Hundley}}, \bibinfo {author} {\bibfnamefont {R.}~\bibnamefont {Movshovich}},
  \bibinfo {author} {\bibfnamefont {J.~L.}\ \bibnamefont {Sarrao}}, \bibinfo
  {author} {\bibfnamefont {J.~D.}\ \bibnamefont {Thompson}}, \bibinfo {author}
  {\bibfnamefont {Z.}~\bibnamefont {Fisk}},\ and\ \bibinfo {author}
  {\bibfnamefont {P.}~\bibnamefont {Monthoux}},\ }\href
  {https://doi.org/10.1088/0953-8984/13/17/103} {\bibfield  {journal} {\bibinfo
   {journal} {Journal of Physics: Condensed Matter}\ }\textbf {\bibinfo
  {volume} {13}},\ \bibinfo {pages} {L337} (\bibinfo {year}
  {2001})}\BibitemShut {NoStop}%
\bibitem [{\citenamefont {Aynajian}\ \emph {et~al.}(2012)\citenamefont
  {Aynajian}, \citenamefont {da~Silva~Neto}, \citenamefont {Gyenis},
  \citenamefont {Baumbach}, \citenamefont {Thompson}, \citenamefont {Fisk},
  \citenamefont {Bauer},\ and\ \citenamefont {Yazdani}}]{Aynajian2012}%
  \BibitemOpen
  \bibfield  {author} {\bibinfo {author} {\bibfnamefont {P.}~\bibnamefont
  {Aynajian}}, \bibinfo {author} {\bibfnamefont {E.~H.}\ \bibnamefont
  {da~Silva~Neto}}, \bibinfo {author} {\bibfnamefont {A.}~\bibnamefont
  {Gyenis}}, \bibinfo {author} {\bibfnamefont {R.~E.}\ \bibnamefont
  {Baumbach}}, \bibinfo {author} {\bibfnamefont {J.~D.}\ \bibnamefont
  {Thompson}}, \bibinfo {author} {\bibfnamefont {Z.}~\bibnamefont {Fisk}},
  \bibinfo {author} {\bibfnamefont {E.~D.}\ \bibnamefont {Bauer}},\ and\
  \bibinfo {author} {\bibfnamefont {A.}~\bibnamefont {Yazdani}},\ }\href
  {https://doi.org/10.1038/nature11204} {\bibfield  {journal} {\bibinfo
  {journal} {Nature}\ }\textbf {\bibinfo {volume} {486}},\ \bibinfo {pages}
  {201} (\bibinfo {year} {2012})}\BibitemShut {NoStop}%
\bibitem [{\citenamefont {Chen}\ \emph {et~al.}(2017)\citenamefont {Chen},
  \citenamefont {Xu}, \citenamefont {Niu}, \citenamefont {Jiang}, \citenamefont
  {Peng}, \citenamefont {Xu}, \citenamefont {Wen}, \citenamefont {Ding},
  \citenamefont {Huang}, \citenamefont {Shu}, \citenamefont {Zhang},
  \citenamefont {Lee}, \citenamefont {Strocov}, \citenamefont {Shi},
  \citenamefont {Bisti}, \citenamefont {Schmitt}, \citenamefont {Huang},
  \citenamefont {Dudin}, \citenamefont {Lai}, \citenamefont {Kirchner},
  \citenamefont {Yuan},\ and\ \citenamefont {Feng}}]{Chen2017}%
  \BibitemOpen
  \bibfield  {author} {\bibinfo {author} {\bibfnamefont {Q.~Y.}\ \bibnamefont
  {Chen}}, \bibinfo {author} {\bibfnamefont {D.~F.}\ \bibnamefont {Xu}},
  \bibinfo {author} {\bibfnamefont {X.~H.}\ \bibnamefont {Niu}}, \bibinfo
  {author} {\bibfnamefont {J.}~\bibnamefont {Jiang}}, \bibinfo {author}
  {\bibfnamefont {R.}~\bibnamefont {Peng}}, \bibinfo {author} {\bibfnamefont
  {H.~C.}\ \bibnamefont {Xu}}, \bibinfo {author} {\bibfnamefont {C.~H.~P.}\
  \bibnamefont {Wen}}, \bibinfo {author} {\bibfnamefont {Z.~F.}\ \bibnamefont
  {Ding}}, \bibinfo {author} {\bibfnamefont {K.}~\bibnamefont {Huang}},
  \bibinfo {author} {\bibfnamefont {L.}~\bibnamefont {Shu}}, \bibinfo {author}
  {\bibfnamefont {Y.~J.}\ \bibnamefont {Zhang}}, \bibinfo {author}
  {\bibfnamefont {H.}~\bibnamefont {Lee}}, \bibinfo {author} {\bibfnamefont
  {V.~N.}\ \bibnamefont {Strocov}}, \bibinfo {author} {\bibfnamefont
  {M.}~\bibnamefont {Shi}}, \bibinfo {author} {\bibfnamefont {F.}~\bibnamefont
  {Bisti}}, \bibinfo {author} {\bibfnamefont {T.}~\bibnamefont {Schmitt}},
  \bibinfo {author} {\bibfnamefont {Y.~B.}\ \bibnamefont {Huang}}, \bibinfo
  {author} {\bibfnamefont {P.}~\bibnamefont {Dudin}}, \bibinfo {author}
  {\bibfnamefont {X.~C.}\ \bibnamefont {Lai}}, \bibinfo {author} {\bibfnamefont
  {S.}~\bibnamefont {Kirchner}}, \bibinfo {author} {\bibfnamefont {H.~Q.}\
  \bibnamefont {Yuan}},\ and\ \bibinfo {author} {\bibfnamefont {D.~L.}\
  \bibnamefont {Feng}},\ }\href {https://doi.org/10.1103/PhysRevB.96.045107}
  {\bibfield  {journal} {\bibinfo  {journal} {Phys. Rev. B}\ }\textbf {\bibinfo
  {volume} {96}},\ \bibinfo {pages} {045107} (\bibinfo {year}
  {2017})}\BibitemShut {NoStop}%
\bibitem [{\citenamefont {Jang}\ \emph {et~al.}(2020)\citenamefont {Jang},
  \citenamefont {Denlinger}, \citenamefont {Allen}, \citenamefont {Zapf},
  \citenamefont {Maple}, \citenamefont {Kim}, \citenamefont {Jang},\ and\
  \citenamefont {Shim}}]{jang2020evolution}%
  \BibitemOpen
  \bibfield  {author} {\bibinfo {author} {\bibfnamefont {S.}~\bibnamefont
  {Jang}}, \bibinfo {author} {\bibfnamefont {J.~D.}\ \bibnamefont {Denlinger}},
  \bibinfo {author} {\bibfnamefont {J.~W.}\ \bibnamefont {Allen}}, \bibinfo
  {author} {\bibfnamefont {V.~S.}\ \bibnamefont {Zapf}}, \bibinfo {author}
  {\bibfnamefont {M.~B.}\ \bibnamefont {Maple}}, \bibinfo {author}
  {\bibfnamefont {J.~N.}\ \bibnamefont {Kim}}, \bibinfo {author} {\bibfnamefont
  {B.~G.}\ \bibnamefont {Jang}},\ and\ \bibinfo {author} {\bibfnamefont
  {J.~H.}\ \bibnamefont {Shim}},\ }\href
  {https://doi.org/10.1073/pnas.2001778117} {\bibfield  {journal} {\bibinfo
  {journal} {Proceedings of the National Academy of Sciences}\ }\textbf
  {\bibinfo {volume} {117}},\ \bibinfo {pages} {23467} (\bibinfo {year}
  {2020})}\BibitemShut {NoStop}%
\bibitem [{\citenamefont {Eo}\ \emph {et~al.}(2021)\citenamefont {Eo},
  \citenamefont {Saha}, \citenamefont {Kim}, \citenamefont {Ran}, \citenamefont
  {Horn}, \citenamefont {Hodovanets}, \citenamefont {Collini}, \citenamefont
  {Fuhrman}, \citenamefont {Nevidomskyy}, \citenamefont {Butch}, \citenamefont
  {Fuhrer},\ and\ \citenamefont {Paglione}}]{Eo2021}%
  \BibitemOpen
  \bibfield  {author} {\bibinfo {author} {\bibfnamefont {Y.~S.}\ \bibnamefont
  {Eo}}, \bibinfo {author} {\bibfnamefont {S.~R.}\ \bibnamefont {Saha}},
  \bibinfo {author} {\bibfnamefont {H.}~\bibnamefont {Kim}}, \bibinfo {author}
  {\bibfnamefont {S.}~\bibnamefont {Ran}}, \bibinfo {author} {\bibfnamefont
  {J.~A.}\ \bibnamefont {Horn}}, \bibinfo {author} {\bibfnamefont
  {H.}~\bibnamefont {Hodovanets}}, \bibinfo {author} {\bibfnamefont
  {J.}~\bibnamefont {Collini}}, \bibinfo {author} {\bibfnamefont {W.~T.}\
  \bibnamefont {Fuhrman}}, \bibinfo {author} {\bibfnamefont {A.~H.}\
  \bibnamefont {Nevidomskyy}}, \bibinfo {author} {\bibfnamefont {N.~P.}\
  \bibnamefont {Butch}}, \bibinfo {author} {\bibfnamefont {M.~S.}\ \bibnamefont
  {Fuhrer}},\ and\ \bibinfo {author} {\bibfnamefont {J.}~\bibnamefont
  {Paglione}},\ }\href@noop {} {\bibfield  {journal} {\bibinfo  {journal}
  {arXiv:2101.03102}\ } (\bibinfo {year} {2021})}\BibitemShut {NoStop}%
\bibitem [{\citenamefont {Fujimori}\ \emph {et~al.}(2019)\citenamefont
  {Fujimori}, \citenamefont {Kawasaki}, \citenamefont {Takeda}, \citenamefont
  {Yamagami}, \citenamefont {Nakamura}, \citenamefont {Homma},\ and\
  \citenamefont {Aoki}}]{Fujimori2019}%
  \BibitemOpen
  \bibfield  {author} {\bibinfo {author} {\bibfnamefont {S.-i.}\ \bibnamefont
  {Fujimori}}, \bibinfo {author} {\bibfnamefont {I.}~\bibnamefont {Kawasaki}},
  \bibinfo {author} {\bibfnamefont {Y.}~\bibnamefont {Takeda}}, \bibinfo
  {author} {\bibfnamefont {H.}~\bibnamefont {Yamagami}}, \bibinfo {author}
  {\bibfnamefont {A.}~\bibnamefont {Nakamura}}, \bibinfo {author}
  {\bibfnamefont {Y.}~\bibnamefont {Homma}},\ and\ \bibinfo {author}
  {\bibfnamefont {D.}~\bibnamefont {Aoki}},\ }\href
  {https://doi.org/10.7566/JPSJ.88.103701} {\bibfield  {journal} {\bibinfo
  {journal} {Journal of the Physical Society of Japan}\ }\textbf {\bibinfo
  {volume} {88}},\ \bibinfo {pages} {103701} (\bibinfo {year}
  {2019})}\BibitemShut {NoStop}%
\bibitem [{\citenamefont {Miao}\ \emph {et~al.}(2020)\citenamefont {Miao},
  \citenamefont {Liu}, \citenamefont {Xu}, \citenamefont {Kotta}, \citenamefont
  {Kang}, \citenamefont {Ran}, \citenamefont {Paglione}, \citenamefont
  {Kotliar}, \citenamefont {Butch}, \citenamefont {Denlinger},\ and\
  \citenamefont {Wray}}]{Miao2020}%
  \BibitemOpen
  \bibfield  {author} {\bibinfo {author} {\bibfnamefont {L.}~\bibnamefont
  {Miao}}, \bibinfo {author} {\bibfnamefont {S.}~\bibnamefont {Liu}}, \bibinfo
  {author} {\bibfnamefont {Y.}~\bibnamefont {Xu}}, \bibinfo {author}
  {\bibfnamefont {E.~C.}\ \bibnamefont {Kotta}}, \bibinfo {author}
  {\bibfnamefont {C.-J.}\ \bibnamefont {Kang}}, \bibinfo {author}
  {\bibfnamefont {S.}~\bibnamefont {Ran}}, \bibinfo {author} {\bibfnamefont
  {J.}~\bibnamefont {Paglione}}, \bibinfo {author} {\bibfnamefont
  {G.}~\bibnamefont {Kotliar}}, \bibinfo {author} {\bibfnamefont {N.~P.}\
  \bibnamefont {Butch}}, \bibinfo {author} {\bibfnamefont {J.~D.}\ \bibnamefont
  {Denlinger}},\ and\ \bibinfo {author} {\bibfnamefont {L.~A.}\ \bibnamefont
  {Wray}},\ }\href {https://doi.org/10.1103/PhysRevLett.124.076401} {\bibfield
  {journal} {\bibinfo  {journal} {Phys. Rev. Lett.}\ }\textbf {\bibinfo
  {volume} {124}},\ \bibinfo {pages} {076401} (\bibinfo {year}
  {2020})}\BibitemShut {NoStop}%
\bibitem [{\citenamefont {Kondo}(1964)}]{kondo1964resistance}%
  \BibitemOpen
  \bibfield  {author} {\bibinfo {author} {\bibfnamefont {J.}~\bibnamefont
  {Kondo}},\ }\href {https://doi.org/10.1143/PTP.32.37} {\bibfield  {journal}
  {\bibinfo  {journal} {Progress of Theoretical Physics}\ }\textbf {\bibinfo
  {volume} {32}},\ \bibinfo {pages} {37} (\bibinfo {year} {1964})}\BibitemShut
  {NoStop}%
\bibitem [{\citenamefont {Choi}\ \emph {et~al.}(2013)\citenamefont {Choi},
  \citenamefont {Haule}, \citenamefont {Kotliar}, \citenamefont {Min},\ and\
  \citenamefont {Shim}}]{choi2013observation}%
  \BibitemOpen
  \bibfield  {author} {\bibinfo {author} {\bibfnamefont {H.~C.}\ \bibnamefont
  {Choi}}, \bibinfo {author} {\bibfnamefont {K.}~\bibnamefont {Haule}},
  \bibinfo {author} {\bibfnamefont {G.}~\bibnamefont {Kotliar}}, \bibinfo
  {author} {\bibfnamefont {B.~I.}\ \bibnamefont {Min}},\ and\ \bibinfo {author}
  {\bibfnamefont {J.~H.}\ \bibnamefont {Shim}},\ }\href
  {https://doi.org/10.1103/PhysRevB.88.125111} {\bibfield  {journal} {\bibinfo
  {journal} {Phys. Rev. B}\ }\textbf {\bibinfo {volume} {88}},\ \bibinfo
  {pages} {125111} (\bibinfo {year} {2013})}\BibitemShut {NoStop}%
\bibitem [{\citenamefont {Choi}\ \emph {et~al.}(2012)\citenamefont {Choi},
  \citenamefont {Min}, \citenamefont {Shim}, \citenamefont {Haule},\ and\
  \citenamefont {Kotliar}}]{choi2012temperature}%
  \BibitemOpen
  \bibfield  {author} {\bibinfo {author} {\bibfnamefont {H.~C.}\ \bibnamefont
  {Choi}}, \bibinfo {author} {\bibfnamefont {B.~I.}\ \bibnamefont {Min}},
  \bibinfo {author} {\bibfnamefont {J.~H.}\ \bibnamefont {Shim}}, \bibinfo
  {author} {\bibfnamefont {K.}~\bibnamefont {Haule}},\ and\ \bibinfo {author}
  {\bibfnamefont {G.}~\bibnamefont {Kotliar}},\ }\href
  {https://doi.org/10.1103/PhysRevLett.108.016402} {\bibfield  {journal}
  {\bibinfo  {journal} {Phys. Rev. Lett.}\ }\textbf {\bibinfo {volume} {108}},\
  \bibinfo {pages} {016402} (\bibinfo {year} {2012})}\BibitemShut {NoStop}%
\bibitem [{\citenamefont {Lu}\ and\ \citenamefont
  {Huang}(2016)}]{lu2016pressure}%
  \BibitemOpen
  \bibfield  {author} {\bibinfo {author} {\bibfnamefont {H.}~\bibnamefont
  {Lu}}\ and\ \bibinfo {author} {\bibfnamefont {L.}~\bibnamefont {Huang}},\
  }\href {https://doi.org/10.1103/PhysRevB.94.075132} {\bibfield  {journal}
  {\bibinfo  {journal} {Phys. Rev. B}\ }\textbf {\bibinfo {volume} {94}},\
  \bibinfo {pages} {075132} (\bibinfo {year} {2016})}\BibitemShut {NoStop}%
\bibitem [{\citenamefont {Kim}\ \emph {et~al.}(2019)\citenamefont {Kim},
  \citenamefont {Ryu}, \citenamefont {Kang}, \citenamefont {Kim}, \citenamefont
  {Choi}, \citenamefont {Nam},\ and\ \citenamefont {Min}}]{kim2019topological}%
  \BibitemOpen
  \bibfield  {author} {\bibinfo {author} {\bibfnamefont {J.}~\bibnamefont
  {Kim}}, \bibinfo {author} {\bibfnamefont {D.-C.}\ \bibnamefont {Ryu}},
  \bibinfo {author} {\bibfnamefont {C.-J.}\ \bibnamefont {Kang}}, \bibinfo
  {author} {\bibfnamefont {K.}~\bibnamefont {Kim}}, \bibinfo {author}
  {\bibfnamefont {H.}~\bibnamefont {Choi}}, \bibinfo {author} {\bibfnamefont
  {T.-S.}\ \bibnamefont {Nam}},\ and\ \bibinfo {author} {\bibfnamefont {B.~I.}\
  \bibnamefont {Min}},\ }\href {https://doi.org/10.1103/PhysRevB.100.195138}
  {\bibfield  {journal} {\bibinfo  {journal} {Phys. Rev. B}\ }\textbf {\bibinfo
  {volume} {100}},\ \bibinfo {pages} {195138} (\bibinfo {year}
  {2019})}\BibitemShut {NoStop}%
\bibitem [{\citenamefont {Zhu}\ \emph {et~al.}(2020)\citenamefont {Zhu},
  \citenamefont {Liu}, \citenamefont {Zhao}, \citenamefont {Wang},
  \citenamefont {Zhang}, \citenamefont {Lu}, \citenamefont {Duan},
  \citenamefont {Xie}, \citenamefont {Feng}, \citenamefont {Jian},
  \citenamefont {Wang}, \citenamefont {Tan}, \citenamefont {Liu}, \citenamefont
  {Zhang}, \citenamefont {Liu}, \citenamefont {Luo}, \citenamefont {Luo},
  \citenamefont {Chen}, \citenamefont {Song},\ and\ \citenamefont
  {Lai}}]{zhu2020kondo}%
  \BibitemOpen
  \bibfield  {author} {\bibinfo {author} {\bibfnamefont {X.-G.}\ \bibnamefont
  {Zhu}}, \bibinfo {author} {\bibfnamefont {Y.}~\bibnamefont {Liu}}, \bibinfo
  {author} {\bibfnamefont {Y.-W.}\ \bibnamefont {Zhao}}, \bibinfo {author}
  {\bibfnamefont {Y.-C.}\ \bibnamefont {Wang}}, \bibinfo {author}
  {\bibfnamefont {Y.}~\bibnamefont {Zhang}}, \bibinfo {author} {\bibfnamefont
  {C.}~\bibnamefont {Lu}}, \bibinfo {author} {\bibfnamefont {Y.}~\bibnamefont
  {Duan}}, \bibinfo {author} {\bibfnamefont {D.-H.}\ \bibnamefont {Xie}},
  \bibinfo {author} {\bibfnamefont {W.}~\bibnamefont {Feng}}, \bibinfo {author}
  {\bibfnamefont {D.}~\bibnamefont {Jian}}, \bibinfo {author} {\bibfnamefont
  {Y.-H.}\ \bibnamefont {Wang}}, \bibinfo {author} {\bibfnamefont {S.-Y.}\
  \bibnamefont {Tan}}, \bibinfo {author} {\bibfnamefont {Q.}~\bibnamefont
  {Liu}}, \bibinfo {author} {\bibfnamefont {W.}~\bibnamefont {Zhang}}, \bibinfo
  {author} {\bibfnamefont {Y.}~\bibnamefont {Liu}}, \bibinfo {author}
  {\bibfnamefont {L.-Z.}\ \bibnamefont {Luo}}, \bibinfo {author} {\bibfnamefont
  {X.-B.}\ \bibnamefont {Luo}}, \bibinfo {author} {\bibfnamefont {Q.-Y.}\
  \bibnamefont {Chen}}, \bibinfo {author} {\bibfnamefont {H.-F.}\ \bibnamefont
  {Song}},\ and\ \bibinfo {author} {\bibfnamefont {X.-C.}\ \bibnamefont
  {Lai}},\ }\href {https://doi.org/10.1038/s41535-020-0248-y} {\bibfield
  {journal} {\bibinfo  {journal} {npj Quantum Materials}\ }\textbf {\bibinfo
  {volume} {5}},\ \bibinfo {pages} {47} (\bibinfo {year} {2020})}\BibitemShut
  {NoStop}%
\bibitem [{\citenamefont {Brito}\ \emph {et~al.}(2018)\citenamefont {Brito},
  \citenamefont {Choi}, \citenamefont {Yao},\ and\ \citenamefont
  {Kotliar}}]{brito2018orbital}%
  \BibitemOpen
  \bibfield  {author} {\bibinfo {author} {\bibfnamefont {W.~H.}\ \bibnamefont
  {Brito}}, \bibinfo {author} {\bibfnamefont {S.}~\bibnamefont {Choi}},
  \bibinfo {author} {\bibfnamefont {Y.~X.}\ \bibnamefont {Yao}},\ and\ \bibinfo
  {author} {\bibfnamefont {G.}~\bibnamefont {Kotliar}},\ }\href
  {https://doi.org/10.1103/PhysRevB.98.035143} {\bibfield  {journal} {\bibinfo
  {journal} {Phys. Rev. B}\ }\textbf {\bibinfo {volume} {98}},\ \bibinfo
  {pages} {035143} (\bibinfo {year} {2018})}\BibitemShut {NoStop}%
\bibitem [{\citenamefont {Zarea}\ \emph {et~al.}(2012)\citenamefont {Zarea},
  \citenamefont {Ulloa},\ and\ \citenamefont {Sandler}}]{zarea2012enhancement}%
  \BibitemOpen
  \bibfield  {author} {\bibinfo {author} {\bibfnamefont {M.}~\bibnamefont
  {Zarea}}, \bibinfo {author} {\bibfnamefont {S.~E.}\ \bibnamefont {Ulloa}},\
  and\ \bibinfo {author} {\bibfnamefont {N.}~\bibnamefont {Sandler}},\ }\href
  {https://doi.org/10.1103/PhysRevLett.108.046601} {\bibfield  {journal}
  {\bibinfo  {journal} {Phys. Rev. Lett.}\ }\textbf {\bibinfo {volume} {108}},\
  \bibinfo {pages} {046601} (\bibinfo {year} {2012})}\BibitemShut {NoStop}%
\bibitem [{\citenamefont {Xu}\ \emph {et~al.}(2019)\citenamefont {Xu},
  \citenamefont {Sheng},\ and\ \citenamefont {Yang}}]{xu2019quasi}%
  \BibitemOpen
  \bibfield  {author} {\bibinfo {author} {\bibfnamefont {Y.}~\bibnamefont
  {Xu}}, \bibinfo {author} {\bibfnamefont {Y.}~\bibnamefont {Sheng}},\ and\
  \bibinfo {author} {\bibfnamefont {Y.-f.}\ \bibnamefont {Yang}},\ }\href
  {https://doi.org/10.1103/PhysRevLett.123.217002} {\bibfield  {journal}
  {\bibinfo  {journal} {Phys. Rev. Lett.}\ }\textbf {\bibinfo {volume} {123}},\
  \bibinfo {pages} {217002} (\bibinfo {year} {2019})}\BibitemShut {NoStop}%
\bibitem [{\citenamefont {Tomczak}(2015)}]{tomczak2015qsgw+}%
  \BibitemOpen
  \bibfield  {author} {\bibinfo {author} {\bibfnamefont {J.~M.}\ \bibnamefont
  {Tomczak}},\ }in\ \href@noop {} {\emph {\bibinfo {booktitle} {Journal of
  Physics: Conference Series}}},\ Vol.\ \bibinfo {volume} {592}\ (\bibinfo
  {organization} {IOP Publishing},\ \bibinfo {year} {2015})\ p.\ \bibinfo
  {pages} {012055}\BibitemShut {NoStop}%
\bibitem [{\citenamefont {Choi}\ \emph {et~al.}(2016)\citenamefont {Choi},
  \citenamefont {Kutepov}, \citenamefont {Haule}, \citenamefont {van
  Schilfgaarde},\ and\ \citenamefont {Kotliar}}]{choi2016first}%
  \BibitemOpen
  \bibfield  {author} {\bibinfo {author} {\bibfnamefont {S.}~\bibnamefont
  {Choi}}, \bibinfo {author} {\bibfnamefont {A.}~\bibnamefont {Kutepov}},
  \bibinfo {author} {\bibfnamefont {K.}~\bibnamefont {Haule}}, \bibinfo
  {author} {\bibfnamefont {M.}~\bibnamefont {van Schilfgaarde}},\ and\ \bibinfo
  {author} {\bibfnamefont {G.}~\bibnamefont {Kotliar}},\ }\href
  {https://doi.org/10.1038/npjquantmats.2016.1} {\bibfield  {journal} {\bibinfo
   {journal} {npj Quantum Materials}\ }\textbf {\bibinfo {volume} {1}},\
  \bibinfo {pages} {16001} (\bibinfo {year} {2016})}\BibitemShut {NoStop}%
\bibitem [{\citenamefont {Choi}\ \emph {et~al.}(2019)\citenamefont {Choi},
  \citenamefont {Semon}, \citenamefont {Kang}, \citenamefont {Kutepov},\ and\
  \citenamefont {Kotliar}}]{choi2019comdmft}%
  \BibitemOpen
  \bibfield  {author} {\bibinfo {author} {\bibfnamefont {S.}~\bibnamefont
  {Choi}}, \bibinfo {author} {\bibfnamefont {P.}~\bibnamefont {Semon}},
  \bibinfo {author} {\bibfnamefont {B.}~\bibnamefont {Kang}}, \bibinfo {author}
  {\bibfnamefont {A.}~\bibnamefont {Kutepov}},\ and\ \bibinfo {author}
  {\bibfnamefont {G.}~\bibnamefont {Kotliar}},\ }\href
  {https://doi.org/https://doi.org/10.1016/j.cpc.2019.07.003} {\bibfield
  {journal} {\bibinfo  {journal} {Computer Physics Communications}\ }\textbf
  {\bibinfo {volume} {244}},\ \bibinfo {pages} {277} (\bibinfo {year}
  {2019})}\BibitemShut {NoStop}%
\bibitem [{\citenamefont {St{\"o}we}(1996)}]{stowe1996contributions}%
  \BibitemOpen
  \bibfield  {author} {\bibinfo {author} {\bibfnamefont {K.}~\bibnamefont
  {St{\"o}we}},\ }\href@noop {} {\bibfield  {journal} {\bibinfo  {journal}
  {Journal of Solid State Chemistry}\ }\textbf {\bibinfo {volume} {127}},\
  \bibinfo {pages} {202} (\bibinfo {year} {1996})}\BibitemShut {NoStop}%
\bibitem [{\citenamefont {Ikeda}\ \emph {et~al.}(2006)\citenamefont {Ikeda},
  \citenamefont {Sakai}, \citenamefont {Aoki}, \citenamefont {Homma},
  \citenamefont {Yamamoto}, \citenamefont {Nakamura}, \citenamefont {Shiokawa},
  \citenamefont {Haga},\ and\ \citenamefont {{\=O}nuki}}]{ikeda2006single}%
  \BibitemOpen
  \bibfield  {author} {\bibinfo {author} {\bibfnamefont {S.}~\bibnamefont
  {Ikeda}}, \bibinfo {author} {\bibfnamefont {H.}~\bibnamefont {Sakai}},
  \bibinfo {author} {\bibfnamefont {D.}~\bibnamefont {Aoki}}, \bibinfo {author}
  {\bibfnamefont {Y.}~\bibnamefont {Homma}}, \bibinfo {author} {\bibfnamefont
  {E.}~\bibnamefont {Yamamoto}}, \bibinfo {author} {\bibfnamefont
  {A.}~\bibnamefont {Nakamura}}, \bibinfo {author} {\bibfnamefont
  {Y.}~\bibnamefont {Shiokawa}}, \bibinfo {author} {\bibfnamefont
  {Y.}~\bibnamefont {Haga}},\ and\ \bibinfo {author} {\bibfnamefont
  {Y.}~\bibnamefont {{\=O}nuki}},\ }\href
  {https://doi.org/10.1143/JPSJS.75S.116} {\bibfield  {journal} {\bibinfo
  {journal} {Journal of the Physical Society of Japan}\ }\textbf {\bibinfo
  {volume} {75}},\ \bibinfo {pages} {116} (\bibinfo {year} {2006})}\BibitemShut
  {NoStop}%
\bibitem [{\citenamefont {Sun}\ and\ \citenamefont
  {Kotliar}(2002)}]{sun2002extended}%
  \BibitemOpen
  \bibfield  {author} {\bibinfo {author} {\bibfnamefont {P.}~\bibnamefont
  {Sun}}\ and\ \bibinfo {author} {\bibfnamefont {G.}~\bibnamefont {Kotliar}},\
  }\href {https://doi.org/10.1103/PhysRevB.66.085120} {\bibfield  {journal}
  {\bibinfo  {journal} {Phys. Rev. B}\ }\textbf {\bibinfo {volume} {66}},\
  \bibinfo {pages} {085120} (\bibinfo {year} {2002})}\BibitemShut {NoStop}%
\bibitem [{\citenamefont {Biermann}\ \emph {et~al.}(2003)\citenamefont
  {Biermann}, \citenamefont {Aryasetiawan},\ and\ \citenamefont
  {Georges}}]{biermann2003first}%
  \BibitemOpen
  \bibfield  {author} {\bibinfo {author} {\bibfnamefont {S.}~\bibnamefont
  {Biermann}}, \bibinfo {author} {\bibfnamefont {F.}~\bibnamefont
  {Aryasetiawan}},\ and\ \bibinfo {author} {\bibfnamefont {A.}~\bibnamefont
  {Georges}},\ }\href {https://doi.org/10.1103/PhysRevLett.90.086402}
  {\bibfield  {journal} {\bibinfo  {journal} {Phys. Rev. Lett.}\ }\textbf
  {\bibinfo {volume} {90}},\ \bibinfo {pages} {086402} (\bibinfo {year}
  {2003})}\BibitemShut {NoStop}%
\bibitem [{\citenamefont {Nilsson}\ \emph {et~al.}(2017)\citenamefont
  {Nilsson}, \citenamefont {Boehnke}, \citenamefont {Werner},\ and\
  \citenamefont {Aryasetiawan}}]{nilsson2017multitier}%
  \BibitemOpen
  \bibfield  {author} {\bibinfo {author} {\bibfnamefont {F.}~\bibnamefont
  {Nilsson}}, \bibinfo {author} {\bibfnamefont {L.}~\bibnamefont {Boehnke}},
  \bibinfo {author} {\bibfnamefont {P.}~\bibnamefont {Werner}},\ and\ \bibinfo
  {author} {\bibfnamefont {F.}~\bibnamefont {Aryasetiawan}},\ }\href
  {https://doi.org/10.1103/PhysRevMaterials.1.043803} {\bibfield  {journal}
  {\bibinfo  {journal} {Phys. Rev. Materials}\ }\textbf {\bibinfo {volume}
  {1}},\ \bibinfo {pages} {043803} (\bibinfo {year} {2017})}\BibitemShut
  {NoStop}%
\bibitem [{\citenamefont {Kutepov}\ \emph {et~al.}(2012)\citenamefont
  {Kutepov}, \citenamefont {Haule}, \citenamefont {Savrasov},\ and\
  \citenamefont {Kotliar}}]{kutepov2012electronic}%
  \BibitemOpen
  \bibfield  {author} {\bibinfo {author} {\bibfnamefont {A.}~\bibnamefont
  {Kutepov}}, \bibinfo {author} {\bibfnamefont {K.}~\bibnamefont {Haule}},
  \bibinfo {author} {\bibfnamefont {S.~Y.}\ \bibnamefont {Savrasov}},\ and\
  \bibinfo {author} {\bibfnamefont {G.}~\bibnamefont {Kotliar}},\ }\href
  {https://doi.org/10.1103/PhysRevB.85.155129} {\bibfield  {journal} {\bibinfo
  {journal} {Phys. Rev. B}\ }\textbf {\bibinfo {volume} {85}},\ \bibinfo
  {pages} {155129} (\bibinfo {year} {2012})}\BibitemShut {NoStop}%
\bibitem [{\citenamefont {Kutepov}\ \emph {et~al.}(2017)\citenamefont
  {Kutepov}, \citenamefont {Oudovenko},\ and\ \citenamefont
  {Kotliar}}]{kutepov2017linearized}%
  \BibitemOpen
  \bibfield  {author} {\bibinfo {author} {\bibfnamefont {A.}~\bibnamefont
  {Kutepov}}, \bibinfo {author} {\bibfnamefont {V.}~\bibnamefont {Oudovenko}},\
  and\ \bibinfo {author} {\bibfnamefont {G.}~\bibnamefont {Kotliar}},\ }\href
  {https://doi.org/https://doi.org/10.1016/j.cpc.2017.06.012} {\bibfield
  {journal} {\bibinfo  {journal} {Computer Physics Communications}\ }\textbf
  {\bibinfo {volume} {219}},\ \bibinfo {pages} {407} (\bibinfo {year}
  {2017})}\BibitemShut {NoStop}%
\bibitem [{\citenamefont {Georges}\ \emph {et~al.}(1996)\citenamefont
  {Georges}, \citenamefont {Kotliar}, \citenamefont {Krauth},\ and\
  \citenamefont {Rozenberg}}]{georges1996dynamical}%
  \BibitemOpen
  \bibfield  {author} {\bibinfo {author} {\bibfnamefont {A.}~\bibnamefont
  {Georges}}, \bibinfo {author} {\bibfnamefont {G.}~\bibnamefont {Kotliar}},
  \bibinfo {author} {\bibfnamefont {W.}~\bibnamefont {Krauth}},\ and\ \bibinfo
  {author} {\bibfnamefont {M.~J.}\ \bibnamefont {Rozenberg}},\ }\href
  {https://doi.org/10.1103/RevModPhys.68.13} {\bibfield  {journal} {\bibinfo
  {journal} {Rev. Mod. Phys.}\ }\textbf {\bibinfo {volume} {68}},\ \bibinfo
  {pages} {13} (\bibinfo {year} {1996})}\BibitemShut {NoStop}%
\bibitem [{\citenamefont {Metzner}\ and\ \citenamefont
  {Vollhardt}(1989)}]{metzner1989correlated}%
  \BibitemOpen
  \bibfield  {author} {\bibinfo {author} {\bibfnamefont {W.}~\bibnamefont
  {Metzner}}\ and\ \bibinfo {author} {\bibfnamefont {D.}~\bibnamefont
  {Vollhardt}},\ }\href {https://doi.org/10.1103/PhysRevLett.62.324} {\bibfield
   {journal} {\bibinfo  {journal} {Phys. Rev. Lett.}\ }\textbf {\bibinfo
  {volume} {62}},\ \bibinfo {pages} {324} (\bibinfo {year} {1989})}\BibitemShut
  {NoStop}%
\bibitem [{\citenamefont {Georges}\ and\ \citenamefont
  {Kotliar}(1992)}]{georges1992hubbard}%
  \BibitemOpen
  \bibfield  {author} {\bibinfo {author} {\bibfnamefont {A.}~\bibnamefont
  {Georges}}\ and\ \bibinfo {author} {\bibfnamefont {G.}~\bibnamefont
  {Kotliar}},\ }\href {https://doi.org/10.1103/PhysRevB.45.6479} {\bibfield
  {journal} {\bibinfo  {journal} {Phys. Rev. B}\ }\textbf {\bibinfo {volume}
  {45}},\ \bibinfo {pages} {6479} (\bibinfo {year} {1992})}\BibitemShut
  {NoStop}%
\bibitem [{\citenamefont {Aryasetiawan}\ \emph {et~al.}(2004)\citenamefont
  {Aryasetiawan}, \citenamefont {Imada}, \citenamefont {Georges}, \citenamefont
  {Kotliar}, \citenamefont {Biermann},\ and\ \citenamefont
  {Lichtenstein}}]{aryasetiawan2004frequency}%
  \BibitemOpen
  \bibfield  {author} {\bibinfo {author} {\bibfnamefont {F.}~\bibnamefont
  {Aryasetiawan}}, \bibinfo {author} {\bibfnamefont {M.}~\bibnamefont {Imada}},
  \bibinfo {author} {\bibfnamefont {A.}~\bibnamefont {Georges}}, \bibinfo
  {author} {\bibfnamefont {G.}~\bibnamefont {Kotliar}}, \bibinfo {author}
  {\bibfnamefont {S.}~\bibnamefont {Biermann}},\ and\ \bibinfo {author}
  {\bibfnamefont {A.~I.}\ \bibnamefont {Lichtenstein}},\ }\href
  {https://doi.org/10.1103/PhysRevB.70.195104} {\bibfield  {journal} {\bibinfo
  {journal} {Phys. Rev. B}\ }\textbf {\bibinfo {volume} {70}},\ \bibinfo
  {pages} {195104} (\bibinfo {year} {2004})}\BibitemShut {NoStop}%
\bibitem [{SM()}]{SM}%
  \BibitemOpen
  \href@noop {} {}\bibinfo {note} {See Supplemental Material for
  details.}\BibitemShut {Stop}%
\bibitem [{\citenamefont {St{\"o}we}(1997)}]{stowe1997uncommon}%
  \BibitemOpen
  \bibfield  {author} {\bibinfo {author} {\bibfnamefont {K.}~\bibnamefont
  {St{\"o}we}},\ }\href@noop {} {\bibfield  {journal} {\bibinfo  {journal}
  {Journal of Alloys and Compounds}\ }\textbf {\bibinfo {volume} {246}},\
  \bibinfo {pages} {111} (\bibinfo {year} {1997})}\BibitemShut {NoStop}%
\bibitem [{\citenamefont {Deng}\ \emph {et~al.}(2019)\citenamefont {Deng},
  \citenamefont {Stadler}, \citenamefont {Haule}, \citenamefont {Weichselbaum},
  \citenamefont {von Delft},\ and\ \citenamefont
  {Kotliar}}]{deng2019signatures}%
  \BibitemOpen
  \bibfield  {author} {\bibinfo {author} {\bibfnamefont {X.}~\bibnamefont
  {Deng}}, \bibinfo {author} {\bibfnamefont {K.~M.}\ \bibnamefont {Stadler}},
  \bibinfo {author} {\bibfnamefont {K.}~\bibnamefont {Haule}}, \bibinfo
  {author} {\bibfnamefont {A.}~\bibnamefont {Weichselbaum}}, \bibinfo {author}
  {\bibfnamefont {J.}~\bibnamefont {von Delft}},\ and\ \bibinfo {author}
  {\bibfnamefont {G.}~\bibnamefont {Kotliar}},\ }\href
  {https://doi.org/10.1038/s41467-019-10257-2} {\bibfield  {journal} {\bibinfo
  {journal} {Nature Communications}\ }\textbf {\bibinfo {volume} {10}},\
  \bibinfo {pages} {2721} (\bibinfo {year} {2019})}\BibitemShut {NoStop}%
\bibitem [{\citenamefont {Koller}\ \emph {et~al.}(2005)\citenamefont {Koller},
  \citenamefont {Hewson},\ and\ \citenamefont {Meyer}}]{koller2005singular}%
  \BibitemOpen
  \bibfield  {author} {\bibinfo {author} {\bibfnamefont {W.}~\bibnamefont
  {Koller}}, \bibinfo {author} {\bibfnamefont {A.~C.}\ \bibnamefont {Hewson}},\
  and\ \bibinfo {author} {\bibfnamefont {D.}~\bibnamefont {Meyer}},\ }\href
  {https://doi.org/10.1103/PhysRevB.72.045117} {\bibfield  {journal} {\bibinfo
  {journal} {Phys. Rev. B}\ }\textbf {\bibinfo {volume} {72}},\ \bibinfo
  {pages} {045117} (\bibinfo {year} {2005})}\BibitemShut {NoStop}%
\bibitem [{\citenamefont {Koga}\ \emph {et~al.}(2005)\citenamefont {Koga},
  \citenamefont {Kawakami}, \citenamefont {Rice},\ and\ \citenamefont
  {Sigrist}}]{koga2005spin}%
  \BibitemOpen
  \bibfield  {author} {\bibinfo {author} {\bibfnamefont {A.}~\bibnamefont
  {Koga}}, \bibinfo {author} {\bibfnamefont {N.}~\bibnamefont {Kawakami}},
  \bibinfo {author} {\bibfnamefont {T.~M.}\ \bibnamefont {Rice}},\ and\
  \bibinfo {author} {\bibfnamefont {M.}~\bibnamefont {Sigrist}},\ }\href
  {https://doi.org/10.1103/PhysRevB.72.045128} {\bibfield  {journal} {\bibinfo
  {journal} {Phys. Rev. B}\ }\textbf {\bibinfo {volume} {72}},\ \bibinfo
  {pages} {045128} (\bibinfo {year} {2005})}\BibitemShut {NoStop}%
\bibitem [{\citenamefont {Lu}\ and\ \citenamefont
  {Huang}(2021)}]{lu2021temperature}%
  \BibitemOpen
  \bibfield  {author} {\bibinfo {author} {\bibfnamefont {H.}~\bibnamefont
  {Lu}}\ and\ \bibinfo {author} {\bibfnamefont {L.}~\bibnamefont {Huang}},\
  }\href {https://doi.org/10.1103/PhysRevB.103.205134} {\bibfield  {journal}
  {\bibinfo  {journal} {Phys. Rev. B}\ }\textbf {\bibinfo {volume} {103}},\
  \bibinfo {pages} {205134} (\bibinfo {year} {2021})}\BibitemShut {NoStop}%
\end{thebibliography}%

\end{document}